\documentclass[11pt,a4paper]{article}
\usepackage{jheppub}
\usepackage{amsmath,amsfonts,amsthm,graphicx,ulem}
\usepackage{float}

\newcommand{\bal}{\begin{align}}
\newcommand{\eal}{\end{align}}
\newcommand{\beq}{\begin{equation}}
\newcommand{\eeq}{\end{equation}}
\newcommand\beqa{\begin{eqnarray}}
\newcommand\eeqa{\end{eqnarray}}
\newcommand\bea{\begin{array}}
\newcommand\eea{\end{array}}

\renewcommand{\leq}{\leqslant}
\renewcommand{\geq}{\geqslant}

    \newcommand{\COMMENT}[1]{}
    
    \newcommand{\neqa}{\nonumber\end{eqnarray}}

\def\[{\left[}
\def\]{\right]}

\def\\langle{\langle}
\def\>{\rangle}

\def\i2{\frac{i}{2}}

\renewcommand{\L}{\mathcal{L}}
\newcommand{\R}{\mathcal{R}}

\title{From Spinning Conformal Blocks \\
to Matrix Calogero-Sutherland Models}
\author[a]{Volker Schomerus,}
\author[b]{Evgeny Sobko}

\affiliation[a]{DESY Hamburg, Theory Group, Notkestra{\ss}e 85, 22607 Hamburg, Germany}
\affiliation[b]{Nordita, Stockholm University and KTH Royal Institute of Technology,
Roslagstullsbacken 23, SE-106 91 Stockholm, Sweden}

\emailAdd{evgenysobko AT gmail.com}
\emailAdd{volker.schomerus AT desy.de}

\abstract{In this paper we develop further the relation between conformal four-point blocks
involving external spinning fields and Calogero-Sutherland quantum mechanics with matrix-valued
potentials. To this end, the analysis of \cite{Schomerus:2016epl} is extended to arbitrary
dimensions and to the case of boundary two-point functions. In particular, we construct the 
potential for any set of external tensor fields.
Some of the resulting Schr\"{o}dinger equations are mapped explicitly to the known Casimir
equations for 4-dimensional seed conformal blocks. Our approach furnishes solutions of
Casimir equations for external fields of arbitrary spin and dimension in terms of functions
on the conformal group. This allows us to reinterpret standard operations on conformal
blocks in terms of group-theoretic objects. In particular, we  shall discuss the relation
between the construction of spinning blocks in any dimension through differential operators 
acting on seed blocks and the action of left/right invariant vector fields on the conformal
group.}

\subheader{DESY-17-177, NORDITA-2017-105}

\usepackage{varioref}
\usepackage{makeidx}
\makeindex
\usepackage{amssymb}

\begin{document}

  \maketitle
\def\SO{\textrm{SO}}
\def\SU{\textrm{SU}}
\def\Spin{\textrm{Spin}}

\section{Introduction}

The theory of conformal partial waves and conformal blocks has a long history that goes
back almost 50 years to early studies in conformal field theory. It was realized from the 
beginning of the subject that conformal partial wave expansions of correlation functions 
provide a very clean way to separate the kinematical skeleton of conformal field theory from 
the dynamical content \cite{Ferrara:1973vz,Mack:1973cwx}. This insight paved the way for the
conformal bootstrap programme \cite{Ferrara:1973yt,Polyakov:1974gs,Mack:1975jr}. Unfortunately,
there was no comprehensive mathematical theory of conformal blocks at the time and we know today 
that even the relevant mathematical background did not yet exist. The entire subject of global  
conformal blocks seemed dormant until the widely recognized papers of Dolan and Osborn 
\cite{Dolan:2000ut,Dolan:2003hv,Dolan:2011dv} which uncovered many highly non-trivial facts about
these functions. With the conception of the modern numerical bootstrap program \cite{Rattazzi:2008pe},
the demands of the scientific community gradually increased. While Dolan and Osborn had focused on
blocks for correlators in which all for external fields are scalar, it is clear that correlators
involving tensor fields, and in particular the stress tensor, provide important additional
constraints. Even though the numerical bootstrap for spinning correlators has only been explored
quite recently, see e.g. \cite{Iliesiu:2015qra,Iliesiu:2017nrv,Dymarsky:2017yzx}, the general
challenge has boosted significant new developments in the theory of spinning conformal blocks
over the last few years, see e.g.\ \cite{Costa:2011dw,Costa:2011mg,SimmonsDuffin:2012uy,
Costa:2014rya,Iliesiu:2015akf,Penedones:2015aga,Rejon-Barrera:2015bpa,Costa:2016xah,Costa:2016hju,
Echeverri:2016dun,Kravchuk:2016qvl,Karateev:2017jgd,Kravchuk:2017dzd}.

There is another area of theoretical physics that was born roughly around the same time as the 
conformal bootstrap,
namely the study of Calogero-Sutherland Hamiltonians. These Hamiltonians were first written
down in \cite{Calogero:1970nt,Sutherland:1971ks,Moser:1975qp} as an interacting multi-particle
(or multi-dimensional) generalization of the famous 1-dimensional Hamiltonian for a single particle
in a P\"oschl-Teller potential \cite{Poschl:1933zz}. The investigation of these models uncovered an
extremely rich structure including spectrum generating symmetries, intriguing dualities and deformations.
Through the seminal work of Heckman and Opdam, see e.g. \cite{Heckman:1987,OpdamDunkl,HeckmanBook} 
and then later of Cherednik, see \cite{Cherednik:2005} and references therein, as well as many others,
Calogero-Sutherland models eventually gave birth to the modern theory of multivariate hypergeometric
functions. Even though a link between Calogero-Sutherland models and group theory had been observed by
Olshanetzki and Perelomov in \cite{Olshanetsky:1976,Olshanetsky:1981,Olshanetsky:1983} and developed further
e.g. in \cite{Etingof:1994,Etingof:1995,Etingof:1993wp}, it was not until
last year that the relevance for conformal blocks and the bootstrap was first pointed out \cite{Isachenkov:2016gim}.
This observation makes an enormous body of recent mathematical results available for conformal field theory
\cite{Isachenkov:2017a}. As we will see below, the application of such techniques goes well beyond scalar
four-point blocks. On the other hand, in the context of spinning blocks, new classes of matrix-valued 
Calogero-Sutherland potentials appear that have received little attention so far. One may therefore 
also hope that some of the methods form the conformal bootstrap could be transferred to the study 
of these new quantum mechanical models. 

The present work lies at the intersection of the two subjects we have sketched above.
In \cite{Schomerus:2016epl} we described a general algorithm to construct certain Calogero-Sutherland
Hamiltonians with matrix-valued potentials for the theory of spinning blocks. For
somewhat technical reasons our general analysis and the examples were restricted to
dimension $d\leq 3$. In the present work we want to overcome this restriction in the 
general case and work out a few examples of relevant matrix Calogero-Sutherland 
models in $d=4$. This will allow us to make contact to the recent work 
\cite{Echeverri:2016dun} on spinning seed blocks in 4-dimensional conformal 
field theory.

The plan of this paper is as follows. In the next section we will review how spinning
conformal blocks can be obtained from vector-valued functions on the conformal group.
Then we explain how the Casimir equation for conformal blocks descends from the
Laplacian on the conformal group and work out an explicit formula that applies to
spinning four-point blocks in any dimension and with any assignment of external
spins. The general expression (\ref{CasEucl}) we derive takes the form of a Schr\"{o}dinger 
operator with some matrix-valued potential that is similar to the one found in
\cite{Feher:2007ooa,Feher:2009wp} for certain quotients of compact groups. This
Hamiltonian acts on functions that depend on cross ratios and take values in
the space of tensor structures. The latter is constructed for the case of 4-dimensional
seed blocks. After insertion into the general expression for the Casimir operator, the
potential takes the form of a matrix-valued Calogero-Sutherland potential. The associated
eigenvalue problem is shown to be equivalent to the Casimir equation of
\cite{Echeverri:2016dun} in Appendix \ref{Apendix4D}. Similarly, we also discuss the case of 
(spinning) boundary two-point functions, derive the corresponding Calogero-Sutherland 
model, and demonstrate the equivalence with the Casimir equations in \cite{Liendo:2012hy}, 
see section 4 and Appendix \ref{ApendixBoundary}. 

Our approach to spinning blocks has several important advantages over conventional ones.
On the one hand, as an approach to Casimir equations it is entirely universal, i.e.
regardless of the setup, whether it involves local spinning fields
as in this work, or is extended to supersymmetric models and defects, the Casimir
equations descend from the Laplacian(s) on the conformal (super)group. This makes many
relations between blocks manifest, see further comments in the concluding section. On
the other hand, our approach realizes solutions of the Casimir equations in terms of
functions on the conformal group itself. In comparison with conventional realizations
of conformal blocks which involve two sets of coordinates, namely coordinates on embedding
space and on an auxiliary space that serves to encode spin degrees of freedom, our realization
treats all variables on the same footing. This throws a new light on the construction
of special spinning blocks from scalar ones \cite{Echeverri:2016dun}, the concept of seed
blocks \cite{Iliesiu:2015qra,Echeverri:2015rwa,Echeverri:2016dun,Costa:2016hju} and
the weight shifting operators that were introduced recently in \cite{Karateev:2017jgd}.
All these possess a simple origin in group theory which we outline in section 5 along 
with some consequences. In particular, we describe a set of seed blocks for conformal 
field theories in any dimension.

\section{The space of conformal blocks}\label{Overview}

In this section we will review the construction of the space of conformal four-point
blocks as a space of equivariant functions on the group. The conformal group in a 
\(d\)-dimensional Euclidean space is given by \(G=\SO(1,d+1)\).
Almost any\footnote{The set of elements that do not have such representation forms
a lower dimensional subspace of zero measure.} element \(h\in G\) admits Bruhat
decomposition:
\begin{gather}
h=\tilde{n}ndr, \ \ \ \tilde{n}\in\tilde{N},\ n\in N, \ d\in D,\ r\in R,
\end{gather}
where \(D=\SO(1,1)\) is a dilatation subgroup, \(R=\SO(d)\) is a subgroup of rotations
and  \(\tilde{N}\), \(N\)  are two abelian subgroups that are generated by translations
and special conformal transformations, respectively.

Finite dimensional irreducible representations of the subgroup $K=D  R \subset G$
are parametrized by the eigenvalue $\Delta$ of the generator of dilatation, also known as
conformal weight, and by the weight $\mu$ of an irreducible representation of the
rotation group. Given these data we can induce an irreducible representation \(\pi_{\Delta,\mu}\)
of the conformal group \(G\). It can be realised on the following space of equivariant functions
\begin{gather}\label{DefPrincSer}
V_{\pi_{\Delta,\mu}} \cong \Gamma^{(\Delta,\mu)}_{G/NDR}=\{ f: G \rightarrow V_\mu  | \ f(hndr)=
e^{\Delta\lambda}\mu(r^{-1})f(h)\}
\end{gather}
and the representation \(\pi_{\Delta,\mu}: G \rightarrow \text{Hom}(V_{\pi_{\Delta,\mu}},V_{\pi_{\Delta,\mu}} )\)
is given by the left regular action of the conformal group 
\begin{gather}\label{LeftAction}
\[\pi_{\Delta,\mu}(h)f\](h')=f(h^{-1}h'), \ \ \ h,h' \in G, \ f\in V_{\pi_{\Delta,\mu}}\ . 
\end{gather}
Here,  $V_{\mu}$ denotes the finite dimensional carrier space of  representation $\mu$
of the rotation group \(R\). We wrote elements $d \in D$ as
\begin{equation}\label{SO(1,1)par}
d(\lambda) = \left( \begin{array}{cc} \cosh\lambda & \sinh\lambda \\
\sinh\lambda & \cosh\lambda \end{array} \right)\ .
\end{equation}
Using the equivariance law we can reconstruct any function $f\in \Gamma^{(\Delta,\mu)}_{G/NDR}$ on
the conformal group $G$ from the values it  assumes on representatives of the $NDR$-orbits. In this
sense, we can also think of the space \eqref{DefPrincSer} as a space of $V_\mu$-valued
functions on the d-dimensional quotient $G/NDR$. Let us add a few comments on the representation
of the rotation group. Throughout the bulk of this work, we will consider the rotations group
SO$(d)$ rather than its universal covering group Spin$(d)$. Correspondingly, the values our label
$\mu$ assumes correspond to representations of SO$(d)$ rather than spinorial representations, i.e.\
we focus on blocks for bosonic external fields. The extension to fermionic fields and spinorial
representations is straight forward. Since our construction of the Calogero-Sutherland Hamiltonian
is local on the conformal group, the distinction between SO$(d)$ and Spin$(d)$ does not matter,
except for the choice of admissible transformations laws $\mu$ of external fields. The latter
can simply assume more values if we want to include external fermions.

The representation (\ref{LeftAction}) belongs to the unitary principal series representations\footnote{There
are also unitary discrete and supplementary series, however they will not appear in our discussion.} of \(G\) if
$\Delta = d/2 + i c$ with $c \in\mathbb{R}$. In what follows we will assume that \(\Delta\) is of this form.
Since all equations for conformal blocks we shall derive below are polynomial in $\Delta$, they can be continued
to arbitrary $\Delta$. In this way, all our equations are applicable to reflection positive Euclidean theories
as well as unitary conformal field theories on a space with Lorentz signature.

We are now prepared to review the construction of four-point blocks from \cite{Schomerus:2016epl}.
Let us pick four external fields which are associated with four representations $\pi_i, i=1, \dots,
4$ of the conformal group. By definition, the space of conformal blocks is given by the space
\((\pi_1\otimes\pi_2\otimes\pi_3\otimes\pi_4)^G\) of $G$-invariants on the four-fold tensor product
\(\pi_1\otimes\pi_2\otimes\pi_3\otimes\pi_4\). In order to construct the space of such invariants
we employ theorem 9.4 of  \cite{Dobrev:1977qv}. It states that  the tensor product of two principal 
series representations  \eqref{DefPrincSer} can be realised on the following space of equivariant 
functions
 \begin{eqnarray}
  \label{TPRC}
  \pi_i \otimes \pi_j & \cong &
  \Gamma^{(\pi_i,\pi_j)}_{G/K} \quad
  \textrm{ with } \\[2mm]  \nonumber
  \Gamma^{(\pi_i,\pi_j)}_{G/K}
 &=&  \left\{
f:G \rightarrow V_{\mu_i} \otimes V_{\mu'_j}\,
\Biggr\rvert  \,
\begin{array}{ll}
 f(hd(\lambda)) = e^{\lambda(\Delta_i-\Delta_j)} f(h) \quad &
\mbox{ for } \ d(\lambda) \in D \subset G \\[2mm]
 f(hr) = \mu_i(r^{-1}) \otimes \mu'_j(r^{-1}) f(h) & \mbox{ for }
r \in R \subset G \end{array} \right\} .
\end{eqnarray}
where  we used the prime symbol for the representation
\(\mu'(r)=\mu(wrw)\) that is twisted by conjugation with the nontrivial element \(w\) of restricted Weyl
group. The latter is given by the quotient \(R'/R\) where \(R'\) is a normalizer of the dilation subgroup
\(D\) within the maximal compact subgroup \(\SO(d+1)\) and it consists of two elements $\{1,w\}$. As a
vector space \(V_{\mu'}\) coincides with \(V_\mu\) and we add the prime in order to stress that we
consider this vector space as a carrier of the representation \(\mu'\).

Just as in our previous discussion of the space \eqref{DefPrincSer} we can also think of
the space \eqref{TPRC} as a space of $V_{\mu_i} \otimes V_{\mu'_j}$-valued function
on the quotient $G/K$. In this case, the underlying quotient space
$G/K$ is $2d$-dimensional. Once again we only used the right regular action of a subgroup
to formulate the equivariance law. Consequently, the conformal group $G$ acts on the space
\eqref{TPRC} through left regular transformations. The associated representation is highly
reducible and may be decomposed into irreducibles of the form \eqref{DefPrincSer}.

In close analogy to eq.\ \eqref{TPRC}, we can realise the two-fold tensor product \(\pi_1\otimes \pi_2\)
on the space of left-equivariant functions,
 \begin{eqnarray}
  \label{TPLC}
  V_{\pi_1 \otimes \pi_2} & \cong &
  \Gamma^{(\pi_1,\pi_2)}_{K\backslash G} \quad
  \textrm{ with } \\[2mm]  \nonumber
  \Gamma^{(\pi_1,\pi_2)}_{K\backslash G}
 &=&  \left\{f:G \rightarrow V_{\mu_1} \otimes V_{\mu'_2}\,
\Biggr\rvert  \,
\begin{array}{ll}
 f(d(\lambda)h) = e^{\lambda(\Delta_2-\Delta_1)} f(h) \quad &
\mbox{ for } \ d(\lambda) \in D \subset G \\[2mm]
 f(rh) = \mu_1(r) \otimes \mu'_2(r) f(h) & \mbox{ for }
r \in R \subset G \end{array} \right\} .
\end{eqnarray}
If we now combine this with the realization of the tensor product  \(\pi_3\otimes \pi_4\) through
right equivariant functions, as in eq.\ (\ref{TPRC}), we arrive at the following realization of the
space \((\pi_1\otimes\pi_2\otimes\pi_3\otimes\pi_4)^G\) of four-point invariants \cite{Schomerus:2016epl},
\begin{equation} \label{sbundles1}
  \Gamma^{(\L\R)}_{K\backslash G/K} = \{\, f:G \rightarrow V_{\L} \otimes
  V_{\R}^\dag \, | \,  f(k_lh k_r^{-1})= \L(k_l)\otimes \R(k_r) f(h) \ , \quad
  \ | \ k_l, k_r \in K \},
\end{equation}
where the two representations \(\L=(a,\mu_1\otimes\mu'_2)\) and \(\R=(-b,\mu_3\otimes\mu'_4)\) of the
subgroup K = \(\SO(1,1)\times \SO(d)\) act on \(V_{\L}= V_{\mu_1} \otimes V'_{\mu_2} \) and \(V_{\R}= V_{\mu_3} \otimes
V_{\mu'_4}\), respectively, according to\footnote{Elements in \(\SO(1,1)\) act on states by multiplication
with a scalar factor. In the following we shall not distinguish between carrier space of representations for
\(\SO(d)\) and \(\SO(1,1)\times \SO(d)\)}
\begin{gather} \label{eq:LRrep}
\L(d(\lambda)r)=e^{2a\lambda}\mu_1(r)\otimes \mu'_2(r)\ \ ,\ \ \
\R(d(\lambda)r)=e^{-2b\lambda}\mu_3(r)\otimes \mu'_4(r)\ .
\end{gather}
Here \(2a=\Delta_2-\Delta_1\), \(2b=\Delta_3-\Delta_4\). Note that the elements $f$ in the space
\eqref{sbundles1} are equivariant with respect to both left and right multiplication of $h$ by
elements $k \in K$. Hence, the space $\Gamma^{(\L\R)}_{K\backslash G/K}$ does no longer admit an
action of the conformal group as one would expect from a space of $G$-invariants.

As in the previous cases, we can use the equivariance laws to reduce a function $f$ to
representatives $a \in A \subset  G$ of the $K \times K$ orbits of $G$. Given the value of
$f$ on any such $a \in A$, we would like to reconstruct $f$ on the entire orbit. But this
might meet an obstacle since the left and right action of $K$ on $G$ may not be independent.
In fact, it turns out that any representative $a$ possesses a nontrivial stabilizer subgroup
in $K \times K$ that is isomorphic to $B=$ SO$(d-2) \subset G$. In order for $f(a)$ to possess
a unique extension to the entire orbit, we should restrict it to take values in the subspace
of \(B\)-invariants \((V_1\otimes V'_2\otimes V_3\otimes V'_4)^B\) which is actually the
space of four-point tensor structures.

Let us explain the statements we made in the previous paragraph in a bit more detail. To
this end we mimic the usual construction of the \(KAK\) or Cartan decomposition and introduce
an automorphism \(\Theta\) acting on \(\xi\in\mathfrak{g}=\text{Lie}(G)\) as \(\Theta(\xi)=
\theta \xi \theta,\ \theta =\, $\textit{diag\/} $(-1,-1,1,\dots,1)\). The map $\Theta$
determines a decomposition of the Lie algebra $\mathfrak{g}$ of the conformal group $G$ as
$\mathfrak{g} = \mathfrak{k} \oplus \mathfrak{p}$ where $\mathfrak{k}=\text{Lie}(K)$ and
$\mathfrak{p}$ its orthogonal complement. To be more concrete let us introduce usual set
of generators $M_{ij}= - M_{ji}$ of the conformal group $G=$$ \SO(1,d+1)$ where $i,j$ run
through $i,j= 0,1,2, \dots d+1$. Obviously, the Lie algebra $\mathfrak{k}$ of $K$ is
spanned by the generator $M_{0,1}$ of dilations along with the elements $M_{\mu\nu}$ for
$\mu = 2, \dots,d+1$ that generate rotations. Our subspace $\mathfrak{p}$ in turn is
spanned by $M_{0,\mu}$ and $M_{1,\mu}$. The subalgebra $\mathfrak{p}$ contains the d-dimensional
abelian subalgebras of translations $P_\mu$ and of special conformal transformations $K_\mu$. We
shall select a 2-dimensional abelian subalgebra $\mathfrak{a}$ that is spanned by $a_+ = M_{0,2}
$ and $a_- = M_{1,3}$. These two generators commute with each other since
they have no index in common. Through exponentiation we pass to the abelian subgroup \(A\subset
\textrm{SO}(1,d+1)\) that consists of matrices of the form
\begin{equation}
\label{adef}
a(\tau_1,\tau_2) =  \left( \begin{array}{ccccccc}
\cosh\frac{\tau_1}{2} & 0 & \sinh\frac{\tau_1}{2} & 0 & 0 & \ldots & 0\\
0 & \cos\frac{\tau_2}{2} & 0 & -\sin\frac{\tau_2}{2} & 0  & \ldots  & 0\\
\sinh\frac{\tau_1}{2} & 0 & \cosh\frac{\tau_1}{2} & 0 & 0 & \ldots & 0\\
0 & \sin\frac{\tau_2}{2} & 0 & \cos\frac{\tau_2}{2} & 0 & \ldots & 0\\
0 & 0 & 0 & 0 & 1 & \ldots & 0\\
\ldots & \ldots & \ldots & \ldots & \ldots & \ldots & \ldots\\
0 & 0 & 0 & 0& 0 & \ldots & 1\\
\end{array} \right), \ \ \ \ a(\tau_1,\tau_2)\in A
\end{equation}
where two variables \(\tau_1 \in (-\infty,\infty)\), \( \tau_2\in [0,4\pi)\) may be considered as 
functions of the two anharmonic ratios one can build from four points in $\mathbb{R}^d$ with $d > 1$. 
Once we know that the space \(A\) coincides with the double
coset \(K\backslash G/K\) we can write the conformal group as  \(G=KAK\). In particular, the
dimension of the space \(KAK\) coincides with that of $G$. Indeed, we can think of \(KAK\) as a
formal union \(KAK= {\cup}_{a \in A} O^{K\times K}_a\)  of \(K\times K\) orbits. Taking into
account that elements $a \in A$ commute with the subgroup \(B=\SO(d-2)\) embedded into the lower
right corner we obtain
 \begin{gather}
\text{dim}(KAK)=2\text{dim}(K)+\text{dim}(A)-\text{dim}(B)=\text{dim}(G)\ .
 \end{gather}
In dimensions $d=2$ and $d=3$ the stabiliser subgroup \(B\) is trivial and \((V_{\L} \otimes V_{\R})^B
=V_{\L} \otimes V_{\R}\). Once we pass to dimensions $d$ higher than $d =3$,  the stabilizer subgroup
\(B=\SO(d-2)\) becomes non-trivial. With our choice (\ref{adef}) of $A$, the subgroup $B \subset K
\subset G$ is embedded into the lower right corner of \((d-2)\times(d-2)\) matrices and such matrices
do commute with all elements $a \in A$ as the latter contain a $(d-2)\times(d-2)$ identity matrix in
the lower-right corner. The action of such elements $b \in B$ on the group $G$ is given through the
embedding $b \mapsto (b,b^{-1}) \in K \times K$ so that an element \(b\in B\) acts on \(a\in A\)
by conjugation \(a\mapsto b a b^{-1} = a\). As we have stressed before, the existence of nontrivial
stabiliser \(B\) leads to ambiguity when we try to reconstruct a function on $G$ from its
values on $A$. Indeed,\ let us take a point \(k_lak_r = k_l b a b^{-1} k_r \in G\). From
the equivariance law in definition (\ref{sbundles1}) conclude that
 \begin{gather}
\L(k_l)\otimes \R(k_r) f(a) =  f(k_lak^{-1}_r)= f(k_lbab^{-1}k^{-1}_r)= \\[2mm]
= \L(k_lb)\otimes \R(k_rb) f(a)
=(\L(k_l)\otimes \R(k_r))(\L(b)\otimes \R(b)) f(a)\ . 
 \end{gather}
In order to ensure that the extension is independent of our choices, we impose that $f$
takes values in the subspace of $B$-invariants. We can construct the projection $\mathcal{P}$
to B-invariants explicitly through
\begin{gather}\label{Projector}
\mathcal{P}=\frac{1}{\text{Vol}(B)}\int\limits_{B} d\mu_b \L(b)\otimes \R(b)\ . 
\end{gather}
This projector is non-trivial for $d > 3$ if at least one of the external primary fields
carries a non-trivial spin. The space \((V_{\L} \otimes V_{\R})^B=\mathcal{P}(V_{\L} \otimes 
V_{\R})\) of $B$-invariants is the image of the projection $\mathcal{P}$.

\section{Calogero-Sutherland Hamiltonian as a radial part of Laplace-Beltrami operator}\label{section3}

\par\medskip
For given intermediate operator of weight $\Delta$ and spin $\mu$ the spinning conformal block
\(\textbf{g}_{\Delta,\mu}\) is a multi-component function of two cross-ratios which diagonalises
the Casimir operator
\begin{gather}
\mathcal{C}^{(2)} [\textbf{g}_{\Delta,\mu}] = C_{\Delta,\mu}\ [\textbf{g}_{\Delta,\mu}]\ ,
\end{gather}
with eigenvalue \(C_{\Delta,\mu}\) equal to the value of quadratic Casimir element in the
representation \(\pi_{\Delta,\mu}\). Of course, higher Casimir elements
are also diagonalized but we will not discuss these any further.

In the previous section we have realized conformal blocks as equivariant functions on the
conformal group which are defined by their restriction to torus \(A\) with two
coordinates \((\tau_1,\tau_2)\) which in turn are functions of anharmonic ratios. In this
section we show that the  Laplace-Beltrami operator, which acts on equivariant functions 
componentwise, descends to the Casimir operator for conformal blocks once it is restricted 
to the torus \(A\).  With an appropriate choice of coordinates
on the conformal group this second order operator
can be worked out explicitly. We have described these coordinates in the previous subsection.
In the first subsection below we will discuss the form of the metric in these coordinates
before we calculate the Casimir operator for spinning blocks in any dimension $d$ in the
second. Our result extends a closely related statement in \cite{Feher:2007ooa} to the case
of conformal groups and the analysis in \cite{Schomerus:2016epl} to $d > 3$. In the  third
subsection we illustrate the general formulas through the example of a spinning seed blocks in
$d=4$ dimensions to recover the Casimir equations that were originally  derived in
\cite{Echeverri:2016dun}.

\subsection{The metric on the conformal group $G$}
\label{ssec:metric} 

As was mentioned in the previous section the space \(G\) is foliated by \((\textit{dim}\ G-2)\)-dimensional
orbits of the \(K\times K\) action on it. These orbits are parametrized by points on the torus
\(A\). We have also stressed that in general the \(K\times K\) action on \(G\) is not free. In fact, 
any given point on the torus is stabilized by the subgroup \(B=\SO(d-2)\).
We can use this gauge freedom by elements $b \in B$ to move e.g. the elements $k_l$ of the left subgroup
$K_l$ to lie in the right coset \(K_l/B\). The corresponding representative will be denoted by $\kappa_l$.
Let us choose some coordinates \(x_l\) on \(K_l/B\) and $(x_r,y)$ on the right subgroup $K_r$ in addition
to the coordinates $\tau=\{\tau_1,\tau_2\}$  on \(A\) that we introduced already. Here, $x_r$ denote
coordinates on the quotient $B\backslash K_r$ and $y$ are coordinates on the subgroup $B$. Hence, points on
\(G\) are parametrized by the coordinates \(\pmb{x}=\{x_l,\tau, x_r,y\}\). We will write $x_A$
for the components of $x$ with $A = 1 , \dots, \textit{dim}\ G$. For the coordinates $x_{l}, x_{r}$ on the
quotients $K_l/B$ and $B\backslash K_r$ we use indices $\alpha = 1, \dots,$ \textit{dim}$K$-\textit{dim}$B$
while coordinates on $B$ are $y_a$ with $a=1, \dots, $\textit{dim}$B$.

In these coordinates we want to compute the metric on the $G$. If we denote an element $h \in G$ by
$h = h(\pmb{x})$, the Killing metric reads
\begin{gather}
  g_{AB}(\pmb{x})=-2\ \text{tr}\ h^{-1}\partial_A h \
  h^{-1}\partial_B h, \ \ \ h\in G \ .
\end{gather}
Let us begin our analysis of the metric with the elements $g_{\alpha i}$ corresponding to a pair of
tangent vectors along $K_l/B$ and $A$, respectively. We parametrize elements \(h\in G\) of conformal
group as \(h=\kappa_l(x_l) a(\tau) k_r(x_r,y)\), where \(\kappa_l\in K_l/B,\ a\in A,\ k_r\in K_r\). A
short evaluation gives
\begin{gather}
  g_{\alpha i}(\pmb{x})=-2\ \text{tr}\ (k_r^{-1}a^{-1}\kappa_l^{-1}\partial_\alpha \kappa_l a k_r)
  (k_r^{-1}a^{-1}\kappa_l^{-1} \kappa_l\partial_i a k_r)=\notag\\[2mm]
  =2\ \text{tr}\ \kappa_l^{-1}\partial_\alpha \kappa_l \partial_i a a^{-1} = 0\ . 
\end{gather}
where the index \(i\) enumerates the coordinates $\tau_i$ on the torus and \(\alpha\) enumerates 
the coordinates \(\{x_{l\alpha}\}\). Here we used orthogonality of the two elements \(\kappa_l^{-1}\partial_\alpha \kappa_l \in \mathfrak{k}\)
and \(\partial_i a \ a^{-1}\in \mathfrak{a}\) with respect to the Killing form. Continuing along these lines
it is easy to see that all elements $g_{Ai} = g_{iA}$ vanish as long as $A \neq i$. Using the explicit
parametrisation (\ref{adef}) for elements of the torus we conclude that the metric tensor has the
following form
\begin{gather}\label{MetricForm}
(g_{AB})=
\begin{pmatrix}
\sharp&...&\sharp&0&0&\sharp&...&\sharp\\
\sharp&...&\sharp&\vdots&\vdots&\sharp&...&\sharp\\
\sharp&...&\sharp&0&0&\sharp&...&\sharp\\
0&...&0&-1&0&0&...&0\\
0&...&0&0&1&0&...&0\\
\sharp&...&\sharp&0&0&\sharp&...&\sharp\\
\sharp&...&\sharp&\vdots&\vdots&\sharp&...&\sharp\\
\sharp&...&\sharp&0&0&\sharp&...&\sharp
\end{pmatrix}\ . 
\end{gather}
This means that the torus \(A\) and any of its images \(k_lAk_r\) under the \(K\times K\)-action crosses
all the \(K\times K\) orbits orthogonally. Generalising terminology of Palais and Terng \cite{Palais:1987} to
the noncompact case, one may therefore call the \(K\times K\)-action \textit{hyperpolar}.

The Haar measure $d\mu_G$ on the conformal group is
$$ d\mu_G = \sqrt{ |\text{det} g|} dx_l \wedge \, d\tau_1 \wedge d\tau_2 \wedge dy \wedge dx_r =
   d\mu_{KaK} \wedge d\tau_1 \wedge d\tau_2 \ , $$
where \(d\mu_{KaK} \) can be interpreted as induced measure on the \(Ka(\tau)K\) orbit. The determinant of
the metric tensor has a factorised form \(\sqrt{ |\text{det} g|}=\omega(\tau_1,\tau_2)\eta(x_l,y,x_r)\) due to
the fact that the $K \times K$ orbits intersect the torus $A$ transversely. The volume
\(\text{vol}(Ka(\tau)K)\) of any given orbit \(Ka(\tau)K\) is given by the integral
\begin{gather}
\text{vol}(Ka(\tau)K)=\int\limits_{Ka(\tau)K}d\mu_{KaK}=\omega(\tau_1,\tau_2)\int dx_l dydx_r\eta(x_l,y,x_r)=
\omega(\tau_1,\tau_2)v_{\infty}\ , 
\end{gather}
where we introduced factor \(v_{\infty}=\int dx_l dydx_r\eta(x_l,y,x_r)\)  which is formally equal to infinity.
This factor is the same for all orbits and can be canceled in all future calculations by the appropriate
normalization described below.

Now we can construct a scalar product on the space of conformal blocks. Let us
first note that the scalar product on the space \eqref{sbundles1} of sections inherits  a
scalar product from the scalar product on the Hilbert space \(L^2_{G}=L^2(G,V_\L \otimes V_\R;
d\mu_{G})\) of vector-valued square integrable functions on the conformal group
\begin{gather}
\int\limits_G d\mu_G(h) \langle f(h),\, g(h)\rangle =  \int\limits_Gd\mu_{G}(\kappa_l,a,k_r)
\langle f(\kappa_l a k_r),\,  g(\kappa_lak_r)\rangle = \notag \\[2mm]
= \int\limits_G d\mu_{G}(\kappa_l,a,k_r)   \langle \L(\kappa_l)\otimes \R(k_r^{-1})f(a),\,
\L(\kappa_l)\otimes \R(k^{-1}_r)g(a)\rangle=\notag\\[2mm]
=\int\limits_G \sqrt{|\text{det}g|}\,  d\pmb{x}\  \langle f(a),\, g(a)\rangle =
v_{\infty}\, \int\limits_A \omega(\tau_1,\tau_2)\  d\tau_1d\tau_2\,
\langle f(a),\,  g(a)\rangle\ ,  \notag
\end{gather}
where \(\langle\dot,\dot\rangle\) is a scalar product in \(V_\L \otimes V_\R\). Taking into account that any
element of the space \eqref{sbundles1} is defined by its restriction \(f(a)\) to \(A\), we obtain
an isomorphism
$$ L^2(\Gamma^{(\L\R)}_{K\backslash G/K};d\mu_G) \ \cong \  L^2_A = L^2(A,(V_\L \otimes
V_\R)^B; d\mu_A)\ . $$
Here, \(d\mu_A=\omega(\tau_1,\tau_2)d\tau_1d\tau_2\) is the measure on the torus $A$ we introduced 
before and the scalar product for the Hilbert space on the right hand side is given by
\begin{gather}
(f(a),g(a))=\mathcal{N}\int\limits_G d\mu_{G}  \langle f(k_lak_r),\,  g(k_lak_r)\rangle=
\int\limits_A d\mu_A \, \langle f(a),\,  g(a)\rangle
\end{gather}
where we introduced normalization factor \(\mathcal{N}={v^{-1}_\infty}\). We can also restrict 
$K \times K$-invariant operators $\mathcal{D}$, such as e.g. the Laplacian, from $L^2_G$ to the 
space $L^2_A$. Matrix elements of  the reduced operators can be computed through the prescription 
\begin{gather}\label{OperReduction}
(f(a),\mathcal{D}^Ag(a))=\mathcal{N}\int\limits_Gd\mu_{G}  \langle \L(\kappa_l)\otimes 
\R(k_r^{-1})f(a),\  \mathcal{D}\  \L(\kappa_l)\otimes \R(k^{-1}_r)g(a)\rangle\ . 
\end{gather}
By construction, the reduced operator $\mathcal{D}_A$ is self-adjoint with respect to the 
scalar product $(\dot,\dot)$ if the original operator $\mathcal{D}$ is.

\subsection{Formula for Laplacian on the torus $A$}

The Laplace-Beltrami operator on any (pseudo)Riemannian manifold is given by the standard
expression
\begin{gather}\label{defLB}
  \Delta_{\textrm{LB}} = \sum\limits_{A,B} |\det (g_{AB})|^{-\frac{1}{2}}
  \partial_{A} g^{A B}|\det (g_{AB})|^{\frac{1}{2}}
  \partial_B \ .
\end{gather}
If we insert the Killing form on the group manifold \(G\) we obtain a Laplace-Beltrami
operator which is invariant under left and right regular transformations (shifts). Initially
the Laplace-Beltrami operator $\Delta_{\textrm{LB}}$ acts on the complex-valued functions. The
extension to $V_\L \otimes V_\R$-valued functions is done by acting componentwise.

Using the definition (\ref{defLB}) and the form (\ref{MetricForm}) of metric tensor we
discussed in the previous subsection, we see that the Laplace-Beltrami operator
\(\Delta_{\textrm{LB}}\) on \(G\) splits into the sum of two terms
\begin{gather}\label{LBasAsum}
\Delta_{\textrm{LB}}=\Delta_{A}+\Delta_{K\times K}\ , 
\end{gather}
where the first term
\begin{gather}
\Delta_A = -\frac{1}{\sqrt{|g|}}\partial_{\tau_1}\sqrt{|g|}\partial_{\tau_1}+
\frac{1}{\sqrt{|g|}}\partial_{\tau_2}\sqrt{|g|}\partial_{\tau_2}
\end{gather}
involves derivatives with respect to the coordinates $\tau_i$ on the torus \(A\) and
the second term contains derivatives with respect to coordinates on $K \times K$ or
rather $K/B \times K$.

With the help of our general prescription (\ref{OperReduction}) we can now reduce the
Laplace-Beltrami operator to sections on the torus \(A\),
\begin{gather}
(f(a),\Delta_{\textrm{LB}}^A\ g(a))=\mathcal{N}\int\limits_G d\mu_{G}  \langle
\L(\kappa_l)\otimes \R(k^{-1}_r)f(a),\, \Delta_{\textrm{LB}}\  \L(\kappa_l)\otimes \R(k^{-1}_r)g(a)\rangle=\notag\\
=\mathcal{N}\int\limits_G d\mu_{KaK}\, d\tau_1 d\tau_2\,   \langle f(a),\,
(\L(\kappa_l)\otimes \R(k^{-1}_r))^{-1} \Delta_{\textrm{LB}}\  \L(\kappa_l)\otimes \R(k^{-1}_r)g(a)\rangle\ , 
\end{gather}
where we used unitarity of the representation \(\L\otimes \R\). Before we continue, let us introduce 
the function
\begin{equation}\label{rho} 
\rho(\kappa_l,k_r) = \L(\kappa_l)\otimes \R(k^{-1}_r)\ .
\end{equation} 
It is defined for a pair of elements $\kappa_l \in K_l/B$ and $k_r \in K_r$ and 
takes values in the linear operators on the finite dimensional vector space $V_\L \otimes V_\R$.  

To proceed with the reduction we
now integrate over the orbits of the $K \times K$ action. For the moment, we want to keep
the choice of the torus $A$ open, i.e. we shall perform our reduction for any $A_{\hat{k}}
\subset G$ that intersects each orbit once. Given $A$, any such $A_{\hat{k}}$ can
be obtained as $A_{\hat{k}} = \hat{\kappa}_{l} A \hat{k}_{r}$ for some choice of $\hat{\kappa}_l, \hat{k}_r
\in K/B \times K$. For the reduction to $A_{\hat{k}}$ we obtain
\begin{gather}
\mathcal{N}\int\limits_Gd\mu_{KaK}\, d\tau_1 d\tau_2\,  \langle f(a),\, \rho(\kappa_l,k_r)^{-1}
\Delta_{\textrm{LB}}\  \rho(\kappa_l,k_r)g(a)\rangle \notag\\
=\int\limits_{A_{\hat{k}}}\omega(\tau_1,\tau_2) d\tau_1d\tau_2 \,  \langle f(a),\, \rho(\hat{\kappa}_{l},
\hat{k}_{r})^{-1} \Delta_{\textrm{LB}}\ \rho(\hat{\kappa}_{l},\hat{k}_{r})g(a)\rangle=\notag\\
=\int\limits_{A}\omega(\tau_1,\tau_2)\, d\tau_1 d\tau_2  \langle f(a),\, \rho(\hat{\kappa}_{l},\hat{k}_{r})^{-1}
\Delta_{\textrm{LB}}\  \rho(\hat{\kappa}_{l},\hat{k}_{r})g(a)\rangle|_{\hat{\kappa}_{l}=1,\ \hat{k}_{r}=1}\ .
\end{gather}
In the course of this short computation we have used the invariance of the measure \(\mu_{KaK}\)
and of the  Laplace-Beltrami operator \(\Delta_{\textrm{LB}}\) under the action of \(K\times K\) on
$G$. The final result is expressed in terms of an integration over our choice $A$
of the torus, i.e.\ we set $\hat{\kappa}_l = 1 = \hat{k}_r$. Note that the original integral over
$G$ is certainly independent of the choice of the torus and hence of $\hat{\kappa}_l$
and $\hat{k}_r$. The function $\rho$ captures the dependence of the reduced Laplace-Beltrami
operator on the choice of the torus.

Using the decomposition (\ref{LBasAsum}), we can evaluate the reduced  the Laplace-Beltrami
operator
 \begin{gather}\label{RedOfLB}
\Delta^A_{\textrm{LB}} = \rho(\hat{\kappa}_l,\hat{k}_r)^{-1} \Delta_{\textrm{LB}}\  \rho(\hat{\kappa}_l,\hat{k}_r)=\notag\\
= -\omega^{-1}\partial_{\tau_1}\omega\partial_{\tau_1}+
\omega^{-1}\partial_{\tau_2}\omega\partial_{\tau_2}+
\rho(\hat{\kappa}_l,\hat{k}_r)^{-1} \Delta_{K\otimes K}\  \rho(\hat{\kappa}_l,\hat{k}_r)\ , 
\end{gather}
where we use \(\omega=\omega(\tau_1,\tau_2)\) to denote the density of the measure on the 
torus $A$ as before. Next we write the second order operator \(\Delta_{K\otimes K}\) 
explicitly in terms of coordinates $(x_r,x_l,y)$ on $K_l/B \times K_r$. If we enumerate 
these coordinates by an index $\mu = 1, \dots, $\textit{dim}$(G)-2$ and denote them as 
$x_\mu$ we find 
\begin{gather}
\rho(\hat{\kappa}_l,\hat{k}_r)^{-1} \Delta_{K\otimes K}\  \rho(\hat{\kappa}_l,\hat{k}_r)=\rho^{-1}
\frac{1}{\sqrt{|g|}}\partial_{x_\mu}(\sqrt{|g|} g^{\mu\nu}\partial_{x_\nu}\rho)=\notag\\
=(\rho^{-1}\partial_{x_\mu}\rho)g^{\mu\nu}(\rho^{-1}\partial_{x_\nu}\rho)+\frac{1}{\sqrt{|g|}}
\partial_{x_\mu}(\sqrt{|g|} g^{\mu\nu}\rho^{-1} \partial_{x_\nu}\rho)\ .
\end{gather}
One may show that the second term vanishes so that the reduced Laplace-Beltrami operator
\(\Delta^A_{\textrm{LB}}\) becomes
\begin{gather}
\Delta^A_{\textrm{LB}}=-\omega^{-1}\partial_{\tau_1}\omega\partial_{\tau_1}+
\omega^{-1} \partial_{\tau_2}\omega\partial_{\tau_2}+(\rho^{-1}
\partial_{x_\mu}\rho)g^{\mu\nu} (\rho^{-1}\partial_{x_\nu}\rho)|_A\ .
\end{gather}
This is almost the result we were after, except that the measure for the integration
over the torus variables $\tau_i$ still involves the nontrivial factor \(\omega\)
which we want to absorb into a redefinition of the functions $f(a)$. So, let us
introduce
\begin{gather}
\psi(\tau_1,\tau_2) = \omega^{\frac{1}{2}} f(\tau_1,\tau_2)\ . 
\end{gather}
On these wave functions, the reduced Laplace-Beltrami operator \(\Delta^A_{\textrm{LB}}\)
acts as
\begin{gather}
H = \omega^{\frac{1}{2}}\ \Delta^A_{\textrm{LB}}\ \omega^{-\frac{1}{2}}\ . 
\end{gather}
In conclusion we have obtained the following final expression for the Casimir operator
in Euclidean  signature
\begin{gather}\label{CasEucl}
H=-\frac{\partial^2}{\partial \tau_1^2}+\frac{\partial^2}{\partial \tau_2^2}-
\omega^{-\frac{1}{2}}(-\frac{\partial^2}{\partial \tau_1^2}+\frac{\partial^2}{\partial \tau_2^2})
\omega^{\frac{1}{2}}+(\rho^{-1}\partial_{x_i}\rho)g^{ij}(\rho^{-1}\partial_{x_j}\rho)|_A\ . 
\end{gather}
It acts on wave functions $\psi$ in the Hilbert space \(L^2(A,V^{\mathcal{P}}, d\tau_1 \wedge d\tau_2)\) with
the canonical scalar product given by
\begin{gather}
(\psi(\tau_1,\tau_2),\phi(\tau_1,\tau_2))=\int\limits_{R\times S^1}d\tau_1d\tau_2\langle \psi(\tau_1,\tau_2),\
\phi(\tau_1,\tau_2)\rangle
\end{gather}
and the vector space \(V^{\mathcal{P}}\) is the image of the four-fold tensor product of the spin
representations under the action of the projection operators $\mathcal{P}$, i.e.\ it is given by
\(V^{\mathcal{P}}=\mathcal{P}\ (V_{\mu_1}\otimes V_{\mu'_2}  \otimes V_{\mu_3} \otimes V_{\mu'_4})\).
We can get the Casimir operator in Lorentz signature by formal analytical continuation with respect
to the coordinates \(x=\frac{1}{2}(\tau_1-i \tau_2)\), \(y=\frac{1}{2}(\tau_1+i \tau_2)\).

In our final expression \eqref{CasEucl}, the Casimir operator takes the form of matrix-valued
Schr\"{o}dinger operator that describes a particle on the strip parametrized by $\tau_1$ and
$\tau_2$. The potential acts in the space $(V_\L \otimes V_\R)^B$ of tensor structures.
It is determined by global features of the foliation of $G$ through $K \times K$
orbits as well as the embedding of the torus $A$ into $G$. In the case of external
scalar fields, i.e. when all the $\mu_i$ are trivial, the potential is found to coincide with
that of a Calogero-Sutherland model for root system $BC_2$ \cite{Isachenkov:2016gim}. In the
more general case of external fields with spin, only one example for $d=3$ has been worked
out before \cite{Schomerus:2016epl}. In this case, the projector $\mathcal{P}$ to SO$(d-2)=$
SO$(1)$-invariant states is trivial. Our task in the final subsection is to illustrate our
 general result \eqref{CasEucl} for the simplest example that involves a non-trivial
projection.

\subsection{Example: Casimir equation for 4-dimensional seed blocks}

In dimension $d=4$ the stabilizer subgroup $B$ is given by \(B=\SO(2)\). In order to illustrate
how our compact formula for the Casimir operator works in this case, we shall study an example
that is relevant for the seed blocks that were introduced in \cite{Echeverri:2016dun}. These
seed blocks occur in the decomposition of correlation functions involving two scalars and two
spinning operators
$$ \langle O_{0,0}O_{s,0}O_{0,0}O_{0,s}\rangle \ , $$
where \(s\in (0,1/2,1,...)\). Two labels \(s_1,s_2\) that we attached to the operators
\(O_{s_1,s_2}\) refer to the representation of the rotation group SO$(4)$. Since the labels
$\mu_1 = (0,0) = \mu_3$ are trivial, our construction the space \eqref{sbundles1} involves
the two representations $\mu_l = \mu_1 \otimes \mu_2' = (s,0)'$ and $\mu_r = \mu_3 \otimes
\mu_4' = (0,s)'$.

Let us now describe the parametrization of $G$ in a bit more detail. In total, we
need $15$ variables. To begin with, we choose coordinates on $K = D \times R$. Using the
isomorphism $r \rightarrow (r_1,r_2)$ between $R = \SO(4)$ and $\SU(2) \times \SU(2)$ (see
appendix \ref{Apendix4D}), our subgroup $K=\textrm{SO}(1,1)\times \textrm{SO}(4) \subset$ SO$(1,5)$ can
be parametrized through three $2\times 2$ matrices of
the form
\begin{gather}
d(\lambda) =\left( \begin{array}{cc} \cosh\lambda & \sinh\lambda \\
\sinh\lambda & \cosh\lambda \end{array} \right)
\end{gather}
\begin{gather}
r_{1}=\begin{pmatrix}
\cos\frac{\theta_{1}}{2}e^{i\frac{\phi_{1}+\psi_{1}}{2}} & i\sin\frac{\theta_{1}}{2}e^{i\frac{\phi_{1}-\psi_{1}}{2}}\\
i\sin\frac{\theta_{1}}{2}e^{-i\frac{\phi_{1}-\psi_{1}}{2}} & \cos\frac{\theta_{1}}{2}e^{-i\frac{\phi_{1}+\psi_{1}}{2}}
\end{pmatrix},\
r_{2} =\begin{pmatrix}
\cos\frac{\theta_{2}}{2}e^{i\frac{\phi_{2}+\psi_{2}}{2}} & i\sin\frac{\theta_{2}}{2}e^{i\frac{\phi_{2}-\psi_{2}}{2}}\\
i\sin\frac{\theta_{2}}{2}e^{-i\frac{\phi_{2}-\psi_{2}}{2}} & \cos\frac{\theta_{2}}{2}e^{-i\frac{\phi_{2}+\psi_{2}}{2}}
\end{pmatrix}\ . 
\end{gather}
When we write an element $h \in G$ as $h = k_l a k_r$, two elements $k_l$ and $k_r$ of $K$
appear. These are parametrized by $\lambda_l, \phi_{li}, \psi_{li}, \theta_{li}$ with $i = 1,2$ and
a similar set of seven variables in which $l$ is replaced by $r$. For the element $a$ of the torus
\(A\) we use the familiar parametrisation (\ref{adef}) in terms of two variables $\tau_1, \tau_2$.
This leaves us with a total number of $16$ variables. Since we are trying to parametrize a
15-dimensional manifold, we need to remove one coordinate. The excess is of course related to
the fact that the group $B = $ SO$(d-2)=$ SO$(2)$ is non-trivial for $d=4$. We can think of $B
\subset R$ as the set of rotation matrices $r = (r_1,r_2)$ for which $\psi_1 = \psi_2$ while all
other parameters vanish, i.e.
\begin{gather}\label{B4d} 
B=\{\, b = (d,r_1,r_2) \in K\,  |\ \lambda_l=\theta_{1}=\theta_{2}=\phi_{1}=\phi_{2}=0 ,\
\psi_{1}=\psi_{2}\} . 
\end{gather}
In the following we shall assume a gauge in which $B$ is removed from $K_l$  by putting \(\psi_{l1}=
-\psi_{l2}=\psi_l\), i.e. we will parametrize the 15-dimensional manifold $\mathcal{G}$ through the
coordinates
$$ \pmb{x} = \{\lambda_l,\theta_{l1},\theta_{l2},\phi_{l1},\phi_{l2}, \psi_{l},\tau_1,\tau_2,
\lambda_r,\theta_{r1},\theta_{r2},\phi_{r1},\phi_{r2},\psi_{r1}, \psi_{r2}\}\ . 
$$
The final technical input that we shall need concerns the form of the generators in the spin-\(s\)
representation of SU$(2)$. These are given by the Wigner matrices \(t^s\), i.e.
their matrix elements take the form
\begin{gather}\label{WignerM}
t^s_{lm}=e^{-i(m\phi+n\psi)}i^{m-n}\left(\frac{(l-m)!(l+m)!}{(l-n)!(l+n)!}\right)^{\frac{1}{2}}
{\sin^{m-n}\theta/2}\cos^{m+n}\theta/2P_{l-m}^{m-n, m+n}(\cos\theta)\ , 
\end{gather}
where \(m\) and \(n\) runs from \(-l\) to \(l\) with integer steps and \(P_n^{a,b}\) is a Jacobi
polynomial. Using this expression one can explicitly build the matrix-valued function \eqref{rho} 
as well as the projector \(\mathcal{P}\).

Let us first demonstrate the general formula \eqref{CasEucl} in the simplest case when \(s=0\), i.e. all
external fields are scalar. This implies that the function $\rho$ receives contributions from the non-trivial
representations of dilations only and is simply given by  \(\rho=\exp(2(a\lambda_l+b\lambda_r))\). If we plug
this simple formula for $\rho$ along with an explicit expression for the metric tensor into formula
\eqref{CasEucl} we obtain the scalar Casimir operator (in Lorentz signature)
\begin{gather}
H_{0}=-\frac{1}{2}\partial_x^2-\frac{1}{2}\partial_y^2+\frac{1}{2}\left( \frac{(a+b)^2-\frac{1}{4}}
{\sinh^2x}-\frac{ab}{\sinh^2\frac{x}{2}}+\frac{(a+b)^2-
\frac{1}{4}}{\sinh^2y}-\frac{ab}{\sinh^2\frac{y}{2}}\right)+\frac{5}{4}\ . 
\end{gather}
As we pointed out at the end of the previous subsection, it takes the form of a 2-particle
Calogero-Sutherland Hamiltonian.

The simplest matrix-valued Hamiltonian appears for \(s=1/2\). With the help of eq.\ (\ref{WignerM}) and
the explicit form for conjugation with the Weyl element \(w\) (see Appendix \ref{Apendix4D})  we
obtain the following expression for the representation matrices
\begin{gather}\label{spin1/2rep}
\mathcal{L}_{\frac{1}{2}}(k_l) \otimes \mathcal{R}_{\frac{1}{2}} (k_r) =
e^{2(a\lambda_l-b\lambda_r)}\times  \\[2mm]
\begin{pmatrix}
\cos\frac{\theta_{l2}}{2}e^{-i\frac{\phi_{l2}+\psi_{l2}}{2}} & i\sin\frac{\theta_{l2}}{2}e^{i\frac{-\phi_{l2}+\psi_{l2}}{2}}\\
i\sin\frac{\theta_{l2}}{2}e^{i\frac{\phi_{l2}-\psi_{l2}}{2}} & \cos\frac{\theta_{l2}}{2}e^{i\frac{\phi_{l2}+\psi_{l2}}{2}}
\end{pmatrix} \otimes
\begin{pmatrix}
\cos\frac{\theta_{r1}}{2}e^{-i\frac{\phi_{r1}+\psi_{r1}}{2}} & i\sin\frac{\theta_{r1}}{2}e^{i\frac{-\phi_{r1}+\psi_{r1}}{2}}\\
i\sin\frac{\theta_{r1}}{2}e^{i\frac{\phi_{r1}-\psi_{r1}}{2}} & \cos\frac{\theta_{r1}}{2}e^{i\frac{\phi_{r1}+\psi_{r1}}{2}}
\end{pmatrix}\ . \notag
\end{gather}
On elements of \(B=\textrm{SO}(2)\), see eq.\ \eqref{B4d}, the map $\rho$  becomes
\begin{gather}
\mathcal{L}_{\frac{1}{2}}(b)\otimes \mathcal{R}_{\frac{1}{2}}(b)=
\begin{pmatrix}
e^{-i\frac{\psi}{2}} & 0 \\
0 & e^{i\frac{\psi}{2}}
\end{pmatrix} \otimes
\begin{pmatrix}
e^{-i\frac{\psi}{2}} & 0 \\
0 & e^{i\frac{\psi}{2}}
\end{pmatrix}=
\begin{pmatrix}
e^{-i\psi} & 0 & 0 & 0 \\
0 & 1 & 0 & 0 \\
0 & 0 & 1 & 0 \\
0 & 0 & 0 & e^{i\psi}
\end{pmatrix}\ . 
\end{gather}
Using our prescription (\ref{Projector}) we obtain the following expression for projector
\begin{gather}\label{Projector1/2}
\mathcal{P}_{\frac{1}{2}}=\frac{1}{4\pi}\int\limits_{0}^{4\pi} d\psi \begin{pmatrix}
e^{-i\psi} & 0 & 0 & 0 \\
0 & 1 & 0 & 0 \\
0 & 0 & 1 & 0 \\
0 & 0 & 0 & e^{i\psi}
\end{pmatrix}=
\begin{pmatrix}
0 & 0 & 0 & 0 \\
0 & 1 & 0 & 0 \\
0 & 0 & 1 & 0 \\
0 & 0 & 0 & 0
\end{pmatrix} \ .
\end{gather}
With the help of eqs. \eqref{spin1/2rep} and \eqref{Projector1/2} we can now evaluate our formula
\eqref{CasEucl} for the Casimir operator. After conjugation with an appropriate constant $2\times 2$
matrix the Hamiltonian $H$ takes the following very simple form
\begin{gather}\label{Spin1halfCasimir}
\begin{pmatrix} \frac{1}{\sqrt{2}} & \frac{1}{\sqrt{2}}\\
-\frac{1}{\sqrt{2}} & \frac{1}{\sqrt{2}}
\end{pmatrix}H_{\frac{1}{2}}\begin{pmatrix}
\frac{1}{\sqrt{2}} & -\frac{1}{\sqrt{2}} \\
\frac{1}{\sqrt{2}} & \frac{1}{\sqrt{2}}
\end{pmatrix}=\begin{pmatrix}
H_0 -\frac{1}{16} & 0 \\
0 & H_0-\frac{1}{16}
\end{pmatrix}+\\[2mm]
\frac{1}{32}\begin{pmatrix}
\frac{1}{\sinh^2\frac{x}{2}}+\frac{1}{\sinh^2\frac{y}{2}}+\frac{4}{\sinh^2\frac{x-y}{4}}-
\frac{4}{\cosh^2\frac{x+y}{4}} & 4(b-a)\left(\frac{1}{\sinh^2\frac{x}{2}}-\frac{1}{\sinh^2\frac{y}{2}}\right) \\
4(b-a) \left( \frac{1}{\sinh^2\frac{x}{2}}-\frac{1}{\sinh^2\frac{y}{2}}\right) & \frac{1}{\sinh^2\frac{x}{2}}+
\frac{1}{\sinh^2\frac{y}{2}}+\frac{4}{\sinh^2\frac{x+y}{4}}-\frac{4}{\cosh^2\frac{x-y}{4}}
\end{pmatrix}\ . \notag
\end{gather}
Similarly, one can derive Hamiltonians for seed blocks with $s \geq 1$ from the Wigner matrix
(\ref{WignerM}) for spin-s representations. The resulting Casimir equations are equivalent to the
ones derived in \cite{Echeverri:2016dun}. For $s=1/2$ we show this equivalence in Appendix \ref{Apendix4D}.
It is interesting to compare the matrix Calogero-Sutherland potential we derived here for 4-dimensional
seed blocks with $s=1/2$ with the matrix potential for fermionic seed blocks in three dimensions we
derived in \cite{Schomerus:2016epl}. The latter was given by a $4\times 4$ matrix that could be
block-diagonalized into two $2\times 2$ matrix potentials. One of these $2\times 2$ blocks for
the 3-dimensional seed blocks has almost the same form as the potential in eq. \eqref{Spin1halfCasimir}
except for slightly different constant pre-factors in front of the interaction terms.

\section{Boundary two-point functions}

Our construction of conformal blocks as functions on the conformal group and of the associated
Casimir equations can be generalised to many others situations. The idea is simple and heuristically
can be formulated as follows: One should split a given correlator on a "left" and a "right" part,
identify their symmetry and then take a double coset over these groups of symmetries. In this way,
all our constructions may be extended to include boundaries, defects and interfaces. Here, we want
to illustrate such extensions at the example of a boundary two-point function.
$$ \langle O_1(x_1) O_2(x_2) \rangle_{Boundary} $$
where the $d-1$-dimensional boundary is assumed to preserve conformal symmetry transformations in
a subgroup SO$(1,d) \subset $ SO$(1,d+1)$. In this case we "split" our system into a left part
containing the boundary and a right part that contains the two local fields and hence is
associated with a tensor product \(\pi_1\otimes\pi_2\) of principal series representations.
The latter can be realized as in eq.\ (\ref{TPRC}) on the right cosets over \(K_r=DR=\textrm{SO}(1,1)
\otimes \textrm{SO}(d)\). On the left side we shall choose the denominator subgroup \(K_l\) to be
the group \(K_l=\textrm{SO}(1,d)\) of conformal transformation in the boundary. In more mathematical
terms, we are instructed to study the following space of functions on the conformal group
\begin{equation} \label{sbundlesBond}
  \Gamma^{(\R)}_{K_l\backslash G/K_r} = \{\, f:\mathcal{G} \rightarrow
  V_{\R} \, | \,  f(k_la k_r^{-1})= \R(k_r) f(a) \  , \forall \ a\in A,\ k_l \in K_l, k_r \in
  K_r \},
\end{equation}
where the representation $\R$ is defined as
\begin{gather} \label{RepRb}
\R(d(\lambda)r)=e^{(\Delta_2-\Delta_1)\lambda}\mu_1(r)\otimes \mu'_2(r)\ .
\end{gather}
Once again we can also think of the space \eqref{sbundlesBond} as a space of sections in a vector
bundle over the double coset \(A=K_l\backslash G/K_r\) with values in the space $V^\mathcal{P} =
(V_1 \otimes V_2)^B$ where the stabilizer group $B$ is now given by $B =$ SO$(d-1)$. In this case,
the double coset turns out to be 1-dimensional and the associated torus can be parametrized as
\begin{equation}
\label{ParAbound}
a(\tau) =  \left( \begin{array}{ccccc}
\cosh\frac{\tau}{2} & 0 & \ldots  & 0 &  \sinh\frac{\tau}{2}\\
0 & 1& \ldots &  0  & 0 \\
\ldots & \ldots & \ldots & \ldots & \ldots \\
0 & 0 & \ldots  & 1 & 0 \\
\sinh\frac{\tau}{2} & 0 & \ldots & 0 & \cosh\frac{\tau}{2}  \\
\end{array} \right)   .
\end{equation}
Following the logic of previous sections we can derive the analogue of the Casimir operator 
in the boundary case. It reads
\begin{gather}\label{CompCasBound}
H=-\frac{\partial^2}{\partial \tau^2}+\omega^{-\frac{1}{2}}\frac{\partial^2}{\partial \tau^2}
\omega^{\frac{1}{2}}+(\rho^{-1}\partial_{x_i}\rho)g^{ij}(\rho^{-1}\partial_{x_i}\rho)|_A\ . 
\end{gather}
where as before \(\omega=\omega(\tau)\) is a factor depending on \(\tau\) in \(\sqrt{|g|}\). 

This Hamiltonian acts on functions $f$ which belong to the Hilbert space \(L^2(A,V^{\mathcal{P}}, 
d \chi)\) with scalar product
\begin{gather}
(f(\tau),g(\tau))=\int\limits_{\mathbb{R}}d\tau\langle f(\tau),\, g(\tau)\rangle
\end{gather}
and the vector space \(V^{\mathcal{P}}\) is defined as the image of the projector
\begin{gather}\label{Bprojector}
\mathcal{P}=\frac{1}{\text{Vol}(B)}\int\limits_{B} d\mu_b \R(b)
\end{gather}
on the carrier space $V_{\mu_1} \otimes V_{\mu'_2}$ of the representation $\R$ we introduced in eq. \eqref{RepRb}.

In case of two scalar operators we have \(\rho=\R(d(\lambda)r)=e^{(\Delta_2-\Delta_1)\lambda}\). Using an explicit
parametrisation of the conformal group $G$ that is adapted to the double coset one can derive the following
Hamiltonian
\begin{gather}
H_d=-\frac{d^2}{d\tau^2}+\frac{1}{16}\left( d^2+ \frac{4(\Delta_1-\Delta_2)^2-1}{\sinh^2\frac{\tau}{2}}-
\frac{(d-3)(d-1)}{\cosh^2\frac{\tau}{2}}\right)
\end{gather}
which can indeed be mapped to the Casimir operator that was derived in \cite{Liendo:2012hy}, see eq.
(\ref{from2012hy}) in Appendix \ref{ApendixBoundary}.
\par\medskip

The generalization to tensor fields in the bulk is now straightforward, following precisely the steps we described
above. As we shall argue in a moment, the corresponding Hamiltonian can be diagonalized into a decoupled systems of
scalar Calogero-Sutherland Hamiltonians in a single variable. To see this, let us pick two representations $(\Delta_1,
\mu_1)$ and $(\Delta_2, \mu_2)$ associated with the two external tensor fields. The tensor product of the finite
dimensional carrier spaces $V_{\mu_1}$ and $V_{\mu'_2}$ can be decomposed with respect to the action of the rotation
group SO$(d)$,
\begin{equation} \label{dec}
 V_{\mu_1}\otimes V_{\mu'_2}=\underset{j}{\oplus}\ V_{\mu_j}\ . 
\end{equation}
The projector $\mathcal{P}$ to SO$(d-1)$ invariant respects this decomposition. When restricted to a single summand
$V_{\mu_j}$ the image of $\mathcal{P}$ is actually 1-dimensional. This is a consequence of the well known fact that
upon restriction from SO$(d)$ to SO$(d-1)$ an irreducible representation of SO$(d)$ decomposes into a sum of
irreducible representations of SO$(d-1)$ with multiplicity at most one. In particular, this is true for the trivial 1-dimensional
representation of SO$(d-1)$ which, if it appears at all, appears a single time. Consequently, the decomposition
\eqref{dec} diagonalizes the Hamiltonian \eqref{CompCasBound} into a set of scalar Calogero-Sutherland, or rather
P\"{o}schl-Teller, models. For $d=3$, for example, these Hamiltonians take the form
\begin{gather}\label{Hd3}
H_{d=3}^s=-\frac{d^2}{d\tau^2}+\frac{9}{16} + \frac{4(\Delta_1-\Delta_2)^2-1}{16\sinh^2\frac{\tau}{2}}-
\frac{s(s+1)}{4\cosh^2\frac{\tau}{2}}\ , 
\end{gather}
where $s = s_j$ is the value the SO$(3)$ spin $s = 0,1,\dots$ takes in the summand $V_{\mu_j}$ of the
decomposition \eqref{dec}. Of course, the results remains true in case we admit fermionic fields for which
representations of the covering group SU$(2)$ of SO$(3)$ appear. If we study the two-point functions for two
fermions, for example, we get two P\"{o}schl-Teller problems of the form \eqref{Hd3} with $s=0,1$.

There is another way to split the setup we discussed here into a ``left'' and a ``right'' system, namely
we can put one of the bulk fields on either side of this split. In this case, since the boundary is included
on both sides, the numerator symmetry is the part of the $d$-dimensional conformal symmetry that is preserved
by the boundary, i.e. the subgroup SO$(1,d) \subset$ SO$(1,d+1)$. Once we add a bulk field, the symmetry is
broken to SO$(d)$. Hence, the relevant double coset for the so-called boundary channel is given by
$$ \textrm{SO}(d)\,  \backslash\, \textrm{SO}(1,d) \, /\,  \textrm{SO}(d) \ . $$
It is easy to see that this double coset is 1-dimensional, i.e. it is parametrized by a single cross ratio.
Working out the Hamiltonians is straightforward. Obviously, one ends up with a set of decoupled P\"{o}schl-Teller
models, one for each SO$(d-1)$ invariant in the tensor product $V_{\mu_1} \otimes V_{\mu_2}$.

\section{Harmonic analysis view on the diagonalization, seed blocks and weight shifting}

The approach we have described in the previous sections implies that all conformal blocks, scalar and
spinning, can be obtained from functions on the conformal group itself. A basis for the latter is well
known from the harmonic analysis of the conformal group. Our main goal in this section is to locate
all spinning blocks within the Hilbert space of functions on the conformal group. This will allow us
to develop a new view on the construction of spinning blocks through differential operators. In
particular we will explain how left and right invariant vector fields can be used to construct arbitrary
functions on the conformal group from a class of seed functions. The construction of certain spinning
blocks from the better studied scalar blocks through differential operators was first advocated in
\cite{Costa:2011dw} and generalized later to a construction of arbitrary spinning blocks from so-called
seed blocks \cite{Echeverri:2015rwa}. In the recent work \cite{Karateev:2017jgd} this development was
carried one step further through so-called weight shifting operators that allow to obtain all spinning
blocks from scalar ones, thereby eliminating more general seed blocks.

According to Peter-Weyl theory, the matrix elements \(\pi_{\Delta,\mu}^{ij}\) of irreducible
unitary representations \(\{\pi_{\Delta,\mu}\}\) form an orthogonal basis in the space of square
integrable functions on the group
\begin{gather}\label{SpectDec}
L^2(G,d\mu_G)=\underset{\Delta,\mu}{\bigoplus} \ \rho_{\pi_{\Delta,\mu}}, \ \ \ \ \
\rho_{\pi_{\Delta,\mu}}=\textrm{span}\{\pi_{\Delta,\mu}^{ij}\} \ .
\end{gather}
Here, the indexes \(i\) and \(j\) enumerate basis vectors in the representation space \(V_{\pi}\).
If $G$ is compact, $\pi$ runs through all irreducible unitary representations. For both compact and non-compact groups such
as the conformal group, however, the precise range of $\pi$ must be determined by solving the spectral
problem for the Laplace-Beltrami operator on the group. In case of conformal group it includes the all unitary principal
series representations\footnote{and also discret series in the case of odd dimension \(d\)}. Namely, the matrix elements of irreducible representations $\pi_{\Delta,\mu}$ 
are eigenfunctions of the scalar Laplace-Beltrami operator,
\begin{gather}
\Delta \pi_{\Delta,\mu}^{ij}(g) = C_{\Delta,\mu} \pi_{\Delta,\mu}^{ij}(g)
\end{gather}
with eigenvalue $C_{\Delta,\mu}$ which is the value of the quadratic Casimir element in the irreducible
representation $\pi_{\Delta,\mu}$. On the other hand, the Hamiltonian (\ref{CasEucl}) for conformal
blocks was obtained as a reduction of Laplace-Beltrami operator with a componentwise action on
vector-valued equivariant functions. Note that components of an eigenfunction $f_{\Delta,\mu}(h)\in
\Gamma^{(\L\R)}_{K\backslash G/K}$ with eigenvalue \(C_{\Delta,\mu}\) can be written as a linear
combination of matrix elements of the representation \(\pi_{\Delta,\mu}\). Upon restriction to the
torus $A$, the equivariant function \(f_{\Delta,\mu}(a)|_{g\in A}\) becomes an eigenfunction of the
Hamiltonian (\ref{CasEucl}). Consequently, after the restriction, the matrix elements
\(\pi^{ij}_{\Delta,\mu}(a(\tau_1,\tau_2))\) provide a basis of functions from which any component
of a general spinning block can be constructed.
\medskip

So far, we have considered the space of square integrable functions as a space of eigenfunctions of the
Laplace operator. But is comes with additional structure. Namely, there exist two commuting actions of the
conformal group $G$ through left and right invariant vector fields. These actions commute with the Laplacian,
i.e. act within the eigenspaces $\rho_{\pi_{\Delta,\mu}}$. Under the combined left and right action, the
eigenspaces are irreducible, i.e. each term in the spectral decomposition \eqref{SpectDec} is a simple
tensor product of irreducible representations for the left and right action of the conformal group on
itself,
\begin{equation} \label{eq:summand}
 L^2_{\Delta,\mu}(G) := \rho_{\pi_{\Delta,\mu}}\cong \Gamma^{\mu;\Delta}_{S^d} \otimes
\Gamma^{\mu^c;\Delta^\ast}_{S^d}\ .
\end{equation}
As before, \(\Gamma^{\mu;\Delta}_{S^d}\) denotes the representation space of the representation
\(\pi_{\Delta,\mu}\) of the conformal group and the second factor in the tensor product is the
conjugate of the first. The lower index \(S^d\) implies that we employ the compact picture for
principal series representation \(\pi_{\Delta,\mu}\) in which vectors are realized as functions
on a d-dimensional sphere \(S^d\), see appendix \ref{CompPicRep} for details. 

As was already mentioned above, components of the spinning conformal blocks can be packaged into
\(L^2(G)\). Let us now describe more precisely, which blocks we can actually construct from functions
in the subspace \(L^2_{\Delta,\mu}(G)\). Here, the compact picture we used in eq.\ \eqref{eq:summand}
is particularly useful since it makes the decomposition of the principal series representations into
irreducible representations of the rotation group quite transparent, see appendix \ref{CompPicRep}. In order
to describe the result, let us write the label $\mu$ of the $R=$ SO$(d)$ representation that defines the space
$\Gamma^{\mu;\Delta}_{S^d}$ of sections in a vector bundle over $S^d$ as $\mu = [k_1, \dots, k_{r-1},k_r]$.
The entries $k_i$ are subject to the restrictions described in eq.\ \eqref{eq:SOdrep}. Then an SO$(d)$
representation $\tilde \mu = [l_1, \dots, l_r]$ appears in the decomposition of
$\Gamma^{\Delta,\mu}_{S^d}$ provided that its labels obey
\begin{equation} \label{eq:condtmumuMain}
 -k_1 \leq l_1 \leq k_2 \quad , \quad k_{i-1} \leq l_i \leq k_{i+1} \quad ,
\quad k_{r-1} \leq l_r \ .
\end{equation}
A derivation of this statement and all the relevant background material on representations of
the rotation group are collected in appendix \ref{CompPicRep}. The way we have presented condition
\eqref{eq:condtmumuMain} it provides us with a list of all the possible covariance laws we can possibly
find within a given eigenspace of the Laplace-Beltrami operator on the conformal group. More precisely,
the covariance law in eq. \eqref{sbundles1} with representations $\mathcal{L}$ and $\mathcal{R}$ as in
eq.\ \eqref{eq:LRrep} can only appear in the sector $L^2_{\Delta,\mu}(G)$  of the tensor products
$\mu_1 \otimes \mu_2'$ and $\mu_3 \otimes \mu_4'$ both contain representations $\tilde \mu_l$ and
$\tilde \mu_r$ satisfying condition \eqref{eq:condtmumuMain}.

As we have stressed above, the eigenspaces $\rho_{\Delta,\mu}$ of the Laplace-Beltrami operator
on the conformal group are irreducible under the combined actin of the first order left- and
right invariant vector fields. This means that we can generate all function in the subspace
$L_{\Delta,\mu}(G)$ from a single seed element by application of differential operators. We
have seen in the previous paragraph that the elements of $L_{\Delta,\mu}(G)$ are associated
with an infinite number of possible covariance laws. There exists a universal choice for the
covariance law of the seed element that works in all sectors and reduces to scalar blocks
in case $\mu = \mu_l$ is a symmetric traceless tensor representation. For a generic choice
of $\mu = [k_1, \dots, k_r]$ , \(r=[d/2]\) the space \eqref{eq:summand} contains an
SO$(d)$ representation with labels
\begin{equation} \label{eq:seedreps}
 \tilde \mu = [l_1=k_2,\dots,l_{r-1} = k_{r-1}, l_r = k_{r-1}]\ .
\end{equation}
Indeed, this set of labels $l_i$ satisfies the condition \eqref{eq:SOdrep} and
hence it describes a representation of SO$(d)$. Moreover, the condition
\eqref{eq:condtmumuMain} is also satisfied which implies that the representation
$\tilde \mu$ of SO$(d)$ appears (with infinite multiplicity) in the decomposition
of the space \eqref{eq:summand} with respect both the left and the right
action of SO$(d)$. Since the space \eqref{eq:summand} is irreducible under the
action of $G_L \times G_R$, any function in $L^2(G)$ can be reconstructed
from those functions of $G$ that satisfy an equivariance condition as in eq.\
\eqref{sbundles1} with $\mathcal{L} =(a,\tilde \mu), \mathcal {R}=(-b,\tilde \mu)$. Hence,
the space of functions on the conformal group $G$ that transform in a
representation of the form \eqref{eq:seedreps} with respect to both left and right
action of the rotation group provide the seeds for the space of all functions. It
allows us to define the set of seed blocks in any dimension \(d\).

Four-point correlators corresponding to the set of seed blocks can be defined in the
following way
\begin{gather}
\langle O_{\Delta_1,\tilde \mu} O_{\Delta_2,0}O_{\Delta_3,{\tilde \mu}^\dag} O_{\Delta_4,0}\rangle\ , 
\end{gather}
where \(\tilde \mu\) is defined as in eq.\ (\ref{eq:seedreps}) and the conformal weights go
over all possible values. For the subspace in which the internal field transforms in a
symmetric traceless tensor representation $\mu_l = [0, \dots, 0,l]$, our choice for
$\tilde \mu$ selects the scalar seed blocks. In the case of $d=4$, which was discussed
in \cite{Echeverri:2016dun}, representations of the rotation group are labeled by two
integers $\mu = [k_1,k_2]$ and we selected representations $\tilde \mu = [k_2,k_2]$ as
seed representations. This agrees with the choice proposed in \cite{Echeverri:2016dun}.
For simplicity, we have restricted our discussion to bosonic fields only, i.e. to
representations of SO$(d)$ rather than Spin$(d)$. But it is easy to generalize everything
we described above to include fermionic representations. In this case, we must also allow
for representation labels in which all entries take half-integer values and upon
restriction from Spin$(d+1)$ to Spin$(d)$ bosonic representations decompose into bosonic
ones while fermionic representations decompose into fermionic ones. With these small
adjustments, our discussion goes through. In the case of $d=3$ one then needs a single
fermionic seed block in addition to the scalar blocks.
\medskip

What we have described so far allows to reconstruct the entire eigenspaces
of the Casimir elements from special eigenfunctions through the application of
left- and right invariant vector fields, or, equivalently, all spinning blocks
from a set of seed blocks through the application of differential operators.
In this sense the construction of blocks is reduced to that of seed blocks.
Quite recently it has been pointed out that there is a simple way to construct
spinning blocks from scalar ones by application of so-called weight shifting
operators. These operators also possess a simple description in the context
of harmonic analysis.

The differential operators we obtained directly from left- and right invariant
vector fields allowed us to construct all blocks with intermediate fields in
symmetric traceless tensor representations $\mu_l$ from scalar blocks, i.e.
we can construct all the subspaces $L^2_{\Delta,\mu_l}(G)$. With our choice
of coordinates on the conformal group $G$, see first paragraph of section
\eqref{ssec:metric}, functions in the subspaces $L^2_{\Delta,\mu_l}(G)$ depend on
variables $x_l,\tau$ and $x_r$, but they are independent of the variables $y$
that parametrize elements in the compact subgroup $B \subset G$. All functions
that do not belong to any $L^2_{\Delta,\mu_l}(G)$ and hence cannot be generated
from scalar blocks by application of left- and right invariant vector fields,
possess a non-trivial dependence on the variables $y$. On the other hand, given
a function that does not depend on $y$, we can actually construct new functions
with non-trivial $y$ dependence by multiplication with those functions that appear
as matrix elements of finite dimensional representations of the conformal group.
These matrix elements do not give rise to normalizable functions on $G$ but if we
multiply such matrix elements with a normalizable functions, the product is
normalizable. In principle it is sufficient to use the matrix elements of the
defining $d+2$-dimensional representation since the latter generate all functions
of the angles $y$. Through iterated multiplication with the $(d+2)^2$ matix
elements of this representation, we can obtain arbitrary functions on the conformal
group from $y$-independent ones, i.e.\ from elements in $L^2_{\Delta,\mu_l}(G)$.
Since the latter may be constructed from scalar blocks by application of left-
and right invariant vector fields, we can build all functions of the group by
the combination of the two constructions we described.

While these methods can give access to formulas for conformal blocks using no more that the
known results for scalar blocks in the sectors $L^2_{\Delta,\mu_l}(G)$, it can become a bit
cumbersome and it gives little control over analytic properties of the blocks as a function
of the various parameters. For these reasons we pursue a different strategy that rests
on the explicit construction of the Casimir equations and their spectrum generating
symmetries, see also comments in the next section.

\section{Discussion}

In this work we have derived a general expression (\ref{CasEucl}) for the Casimir equation of spinning
four-point blocks in any dimension in terms of some Schr\"{o}dinger problem. The universality
of our approach is one of its main advantages since it makes standard mathematical tools
applicable for the study of a wide variety of conformal blocks, involving external tensor
fields, boundaries and more general defects. Many of these tools will be discussed at the
example of scalar four-point functions in the forthcoming papers \cite{Isachenkov:2017a}
and \cite{Isachenkov:2017b}. There are many interesting further directions to pursue.

On the one hand, it is certainly important to extend the analysis in section \ref{section3} and to
derive more explicit expressions for the potentials. As a first important step towards a
systematic solution theory one should then analyse the symmetries and singularities of
these potentials. Next one needs to investigate the behaviour of solutions in the
vicinities of the singularities in order to understand the precise analytic structure
of solutions (blocks) in the space of cross ratios. The Calogero-Sutherland form of the
Casimir equation is also well suited to derive series expansions, recurrence relations
and to study the analytic properties of solutions in the space of eigenvalues, which
include the weight $\Delta$ of the intermediate fields as well as the spin $\mu$. 
Furthermore, it would be interesting to study
(super-)integrability and spectrum generating symmetries of our matrix-valued
Calogero-Sutherland models. It should be possible to develop all these issues in close
analogy to the scalar case, see \cite{Isachenkov:2017a,Isachenkov:2017b}.

The coset spaces we met in the context of conformal blocks lead to Calogero-Sutherland
Hamiltonians of type \textit{BC}$_N$. This means that the terms in the scalar potential
are in one-to-one correspondence with the positive roots on a \textit{BC}$_N$ root
system. Calogero-Sutherland systems associated with \textit{A}$_N$ root systems have
been studied more extensively. These arise in the context of coset spaces $G/G$ where the
denominator groups acts by conjugation. In \cite{Reshetikhin:2015pma,Reshetikhin:2015rba}
Reshetikhin introduced a matrix version of these $A_N$ Calogero-Sutehrland models that is
very similar to the matrix \textit{BC}$_N$ models we constructed above. For the $A_N$
series these matrix systems were shown to be super (or degenerate) integrable. Is
therefore seems likely that the same is true for the matrix Calogero-Sutherland
Hamiltonians that emerge from the theory of spinning conformal blocks.

As we have stressed several times, our approach to Casimir equations and conformal
blocks is very flexible. In particular it also applies to superblocks, including
cases involving external fields do not belong to BPS multiplets, which has not
received much attention yet. If the external fields sit in half-BPS multiplets, the
denominator subgroups $K_l$ and $K_r$ contain enough fermionic generators to remove
all fermionic directions from the double coset. The associated Calogero-Sutherland
models are purely bosonic and look similar to the ones we discussed above. The
opposite case in which all external fields belong to long multiplets has only been
studied in 2 dimensions, see \cite{Fitzpatrick:2014oza} for the case of
$\mathcal{N}=1$ superconformal symmetry and \cite{Cornagliotto:2017dup} for
$\mathcal{N}=2$. In the latter paper clear evidence was given that the superblock
decomposition of correlation functions for primaries in long multiplets is more
constraining than the decomposition into bosonic blocks or the study of BPS
correlators. It therefore seems worthwhile to develop a supersymmetric version of
the approach we described above, construct the associated super Calogero-Sutherland
models and develop a systematic theory of superblocks.

A final line of applications we want to mention concerns the study of blocks in the
presence of boundaries or defects \cite{Liendo:2012hy,Gadde:2016fbj,Billo:2016cpy,
Liendo:2016ymz,Fukuda:2017cup}. All these possess a description in terms of cosets
of the conformal group and the Casimir equations take the form of a Calogero-Sutherland
Hamitonian \cite{Schomerus:2017}. Furthermore, our approach might be useful for the
application of bootstrap methods to correlation functions of particular nonlocal
operators such as the BFKL light-ray operators introduced in \cite{Balitsky:2013npa}.
For these operators, the operator product coefficients were already calculated in
\cite{Balitsky:2015tca,Balitsky:2015oux}. We will come back to these and related
issues in future research.

\begin{acknowledgments}
\label{sec:acknowledgments}
We thank Misha Isachenkov, Edwin Langmann, Pedro Liendo, Madalena Lemos and Kostya Zarembo
for interesting discussions. The work of E. Sobko was supported by the grant "Exact Results
in Gauge and String Theories" from the Knut and Alice Wallenberg foundation and, in part,
by the ERC advanced grant No 341222.
\end{acknowledgments}

\appendix

\section{Comparison with Casimir equations for 4d seed blocks}\label{Apendix4D}

In this appendix we want to compare the Calogero-Sutherland Hamiltonian for the $s=1/2$ seed block
we have constructed in eq.\ \eqref{Spin1halfCasimir} with the Casimir equations that were put forward
in \cite{Echeverri:2016dun}. After a bit of background on the relevant group theory, we shall review
that results form \cite{Echeverri:2016dun} for the $s=1/2$ Casimir equations. Then we describe in
detail how to map these Casimir equations to the Schr\"{o}dinger equation for the matrix Calogero-Sutherland
model \eqref{Spin1halfCasimir}.

Let us begin with a few comments on the well known isomorphism between Spin$(4)$ and $\textrm{SU}(2)\times
\textrm{SU}(2)$. As usual the first step is to identify quaternions with \(2\times 2\)-matrices as
\begin{gather}
q=a+ib+jc+kd \ \leftrightarrow\  \begin{pmatrix}
a+ib & c+id \\
-c+id & a-ib
\end{pmatrix}\ . 
\end{gather}
Then we introduce the group of left and right unit quaternions which is isomorphic to \(\SU(2)\)
 \begin{gather}
 q_1\leftrightarrow \begin{pmatrix}
\cos\frac{\theta_1}{2}e^{i\frac{\phi_1+\psi_1}{2}} & i\sin\frac{\theta_1}{2}e^{i\frac{\phi_1-\psi_1}{2}}\\
i\sin\frac{\theta_1}{2}e^{-i\frac{\phi_1-\psi_1}{2}} & \cos\frac{\theta_1}{2}e^{-i\frac{\phi_1+\psi_1}{2}}
\end{pmatrix},\
 q_2\leftrightarrow \begin{pmatrix}
\cos\frac{\theta_2}{2}e^{i\frac{\phi_2+\psi_2}{2}} & i\sin\frac{\theta_2}{2}e^{i\frac{\phi_2-\psi_2}{2}}\\
i\sin\frac{\theta_2}{2}e^{-i\frac{\phi_2-\psi_2}{2}} & \cos\frac{\theta_2}{2}e^{-i\frac{\phi_2+\psi_2}{2}}
\end{pmatrix}\ . 
\end{gather}
Any pair \((q_1,q_2)\in \SU(2)\times \SU(2)\) of such elements acts on $\mathbb{R}^4$ through
\begin{gather}
(q_1,q_2): \mathbb{R}^4 \, \mapsto \, \mathbb{R}^4 \quad  (q_1,q_2):x \, \rightarrow \, q_1xq_2^{-1}\ .
\end{gather}
On the right hand side we have represented the elements $x \in \mathbb{R}^4$ through the associated
element in the space of quaternions,
\begin{gather}
x \  \leftrightarrow\ \begin{pmatrix}
x_1+ix_2 & x_3+ix_4 \\
-x_3+ix_4 & x_1-ix_2
\end{pmatrix}\ . 
\end{gather}
This action of $(q_1,q_2)$ on $\mathbb{R}^4$ allows us to assign an element of $\SO(4)$ to any element
$(q_1,q_2) \in \SU(2) \times \SU(2)$. The map is well known to be an isomorphism, i.e.\ any rotation of
$\mathbb{R}^4$ is of that form.
\par\medskip
There is one more piece of information about the group theory we used in deriving our Calogero-Sutherland
Hamiltonian \eqref{Spin1halfCasimir}, namely the non-trivial element $w$ of the restricted Weyl group. It
can be represented as a $6 \times 6$ matrix \(w=\text{diag}\{1,-1,1,1,1,-1\}\) which acts on the 4-dimensional
conformal group $\SO(1,5)$ by conjugation. When restricted to elements \(r\in \SO(4)\subset \SO(1,5)\) the
actions reads
\begin{gather}
wr(\theta_1,\phi_1,\psi_1,\theta_2,\phi_2,\psi_2)w=r(\theta_2,-\phi_2,-\psi_2,\theta_1,-\phi_1,-\psi_1)\ . 
\end{gather}
This concludes our brief discussion of group theoretic background.

Let us now turn to checking the equivalence between Hamiltonian (\ref{Spin1halfCasimir}) and  Casimir operator
derived in \cite{Iliesiu:2015akf}. In the case of \(s=1/2\), the equations (3.17) from \cite{Iliesiu:2015akf}
read as
\begin{gather}
(\Delta_3^{a+\frac{1}{4},b+\frac{3}{4};1}-\frac{1}{8})G^{(1)}_0+\frac{1}{2}L(b-\frac{1}{4})G_1^{(1)}=
\frac{E^1_l}{2}G^{(1)}_0 \label{1stEqn} \\[2mm]
2z\bar{z}L(a+\frac{1}{4})G_0^{(1)}+(\Delta_3^{a+\frac{1}{4},b-\frac{1}{4};0}+\frac{15}{8})G_1^{(1)}=
\frac{E^1_l}{2}G^{(1)}_1\label{2ndEqn}
\end{gather}
where
\begin{gather}
\Delta_\epsilon^{(a,b;c)}=D_z^{(a,b;c)}+D_{\bar{z}}^{(a,b;c)}+\epsilon \frac{z\bar{z}}{z-\bar{z}}
\left((1-z)\partial_z-(1-\bar{z})\partial_{\bar{z}}\right),\\
D_z^{(a,b;c)}=z^2(1-z)\partial^2_z-((a+b+1)z^2-cz)\partial_z-abz
\end{gather}
and
\begin{gather}
L(\mu)=-\frac{1}{z-\bar{z}}\left(z(1-z)\partial_z-\bar{z}(1-\bar{z})\partial_{\bar{z}}\right)+\mu\ . 
\end{gather}
Introducing new variables \(z=-\sinh^{-2}\frac{x}{2}\), \(\bar{z}=- \sinh^{-2}\frac{y}{2}\) we
can rewrite eqs.\ (\ref{1stEqn}) - (\ref{2ndEqn}) in the form of eigenvalue problem for differential
matrix operator
\(\tilde{\mathcal{M}}\)
\begin{gather}
\tilde{\mathcal{M}}\begin{pmatrix}
G_0^{(1)}  \\
G_1^{(1)}
\end{pmatrix}=\frac{E_l^1}{2}\begin{pmatrix}
G_0^{(1)}  \\
G_1^{(1)}
\end{pmatrix}\ . 
\end{gather}
We will not write out the precise form of $\tilde{\mathcal M}$. It is easily obtained from the formulation
in eqs.\ \eqref{1stEqn} and \eqref{2ndEqn}. Next one needs to perform a gauge transformation of the form
\begin{gather}
\mathcal{M}=S^{-1}\tilde{\mathcal{M}}S\ , 
\end{gather}
where the matrix $S$ is given by
\begin{gather}
S=\begin{pmatrix}
\chi_0 & 0 \\
0  & 1
\end{pmatrix}
\begin{pmatrix}
-1 & 1 \\
1  & 1
\end{pmatrix}
\begin{pmatrix}
\chi_1 & 0 \\
0  & \chi_2
\end{pmatrix}
\begin{pmatrix}
0 & 1 \\
1  & 0
\end{pmatrix},\\[2mm]
\chi_0=\frac{1}{2}\sinh\frac{x}{2}\sinh\frac{y}{2},\notag\\[2mm]
\chi_1=\frac{(\cosh\frac{x}{2}\cosh\frac{y}{2})^{-\frac{1}{2}-a-b}(\sinh\frac{x}{2}\sinh\frac{y}{2})^{-\frac{1}{2}+a+b}}
{(\sinh\frac{x}{2}-\sinh\frac{y}{2})(\sinh\frac{x}{2}+\sinh\frac{y}{2})^2},\notag \\[2mm]
\chi_2=\frac{(\cosh\frac{x}{2}\cosh\frac{y}{2})^{-\frac{1}{2}-a-b}(\sinh\frac{x}{2}\sinh\frac{y}{2})^{-\frac{1}{2}+a+b}}
{(\sinh\frac{x}{2}-\sinh\frac{y}{2})^2(\sinh\frac{x}{2}+\sinh\frac{y}{2})} \ . \notag
\end{gather}
A short and explicit calculation shows that \(\mathcal{M}\)  and our Hamiltonian (\ref{Spin1halfCasimir})
are related by
\begin{gather}
\mathcal{M}=-2\begin{pmatrix}
\frac{1}{\sqrt{2}} & \frac{1}{\sqrt{2}} \\
-\frac{1}{\sqrt{2}} & \frac{1}{\sqrt{2}}
\end{pmatrix}H_{\frac{1}{2}}\begin{pmatrix}
\frac{1}{\sqrt{2}} & -\frac{1}{\sqrt{2}} \\
\frac{1}{\sqrt{2}} & \frac{1}{\sqrt{2}}\ .
\end{pmatrix}
\end{gather}
While our explicit analysis of the matrix Calogero-Sutherland Hamiltonian for 4-dimensional seed blocks and its
comparison with the Casimir equations in \cite{Echeverri:2016dun} was restricted to the first non-trivial case
$s=1/2$, it is clear that similar results hold for all the other Casimir equations.

\section{Comparison with Casimir equations for boundary two-point function}\label{ApendixBoundary}

The goal of this appendix is to compare the P\"{o}schl-Teller Hamiltonian \eqref{CompCasBound} that we
obtained in our discussion of two-point functions of scalar fields in the presence of a boundary with
the Casimir operator derived in \cite{Liendo:2012hy}. The correlation function of two scalar operators
in the presence of a boundary has the following general form \cite{Liendo:2012hy}
\begin{gather}\label{2ScB}
\langle O_1(x_1) O_2(x_2) \rangle= \frac{1}{(2x_1^d)^{\Delta_1}(2 x_2^d)^{\Delta_2}}\xi^{-(\Delta_1+\Delta_2)/2}G(\xi)\ , 
\end{gather}
where \(\xi=\frac{(x_1-x_2)^2}{4x_1^d x_2^d}\) is the anharmonic ratio. Equation (A.4) from \cite{Liendo:2012hy}
reads
\begin{gather} \label{PLeq}
\xi(1+\xi)g''(\xi)+(\Delta-\frac{d}{2}+1+(\Delta+1)\xi)g'(\xi)+\frac{1}{4}(\Delta^2-(\Delta_1-\Delta_2)^2)g(\xi)=0\ , 
\end{gather}
where \(g(\xi)=\xi^{-\frac{\Delta}{2}}G(\xi)\) and \(G(\xi)\) is defined in eq.\ (\ref{2ScB}). Introducing the new
variable \(\xi=\sinh^{-2}\frac{\tau}{2}\) we can rewrite eq.\ \eqref{PLeq} in the form
\begin{gather}
\tilde{H}_dG(\tau)=\frac{\Delta(d-\Delta)}{4} G(\tau),
\end{gather}
where
\begin{gather}
\tilde{H}_d=-\frac{d^2}{d\tau^2}-(1-\frac{d}{2}(1-\cosh\tau))
\frac{1}{\sinh \tau} \frac{d}{d\tau}+\frac{(\Delta_1-\Delta_2)^2}{\sinh^2\frac{\tau}{2}}\ .
\end{gather}
Through a gauge transformation with the scalar function
$$\chi(\tau)=\left(\frac{\sinh\frac{\tau}{2}}{\cosh\frac{\tau}{2}}\right)^{\frac{d-2}{4}}
\frac{1}{\sinh^{\frac{d}{4}}\tau}$$
brings the operator $\tilde{H}_d$ into the form
\begin{gather}
H_d=\chi^{-1}(\tau)\circ \tilde{H}_d\circ \chi(\tau)\notag\\
=-\frac{d^2}{d\tau^2}+\frac{1}{16}\left( d^2+ \frac{4(\Delta_1-\Delta_2)^2-1}
{\sinh^2\frac{\tau}{2}}-\frac{(d-3)(d-1)}{\cosh^2\frac{\tau}{2}}\right) \label{from2012hy}
\end{gather}
which agrees with the P\"{o}schl-Teller Hamiltonian we found in section 4. This concludes our
comparison with the results from \cite{Liendo:2012hy}.

\section{Compact picture for principal series representations of $G$}\label{CompPicRep}

In this appendix we describe how the principal
continuous series decompose with respect to the subgroup $R \subset K \subset
G$.  It turns out that this is most easily described if we  pass to a
model for the principal series representation $\pi_{\Delta,\mu}$ in which the carrier space
is realized in terms of functions on the quotient $S^{d} =$ SO$(d+1)$/SO$(d)$ of the maximally
compact subgroup $L = SO(d+1)$. More precisely, it is known \cite{Dobrev:1977qv} that we can realize
$\pi_{\Delta,\mu}$ on the space
$$ \Gamma^{\Delta,\mu}_{G/NDR} \cong \Gamma^{\Delta,\mu}_{S^d} = \{ f: \textrm{SO}(d+1)
\rightarrow V_\mu\, | \, f(ur) = \mu(r^{-1}) f(k) \, \} \ . $$
Note that this space does not depend on the choice of $\Delta$. The dependence on
$\Delta$ comes in when we introduce the action of $G$ on $\Gamma^{\Delta,\mu}_{S^d}$ which
is given by
$$ \pi_{\Delta,\mu}(g) f(u) = e^{\Delta\lambda} f(u_g)   \quad \textrm{where} \quad
g^{-1}u = u_g n d(\lambda)\  $$
for $u,u_g \in$ SO$(d+1)$. Obviously, the space $\Gamma^{\Delta,\mu}_{S^d}$ carries
a representation of the isometry group $L =$ SO$(d+1)$ of the sphere. It will be important
for us to understand how the space it decomposes under this action, but before we can spell
out a precise statement, we need a bit of background on representations of orthogonal groups.

For simplicity we shall again assume that $d=2r$ is even. The odd case can be treated
similarly. Let us recall that finite dimensional representations $\varrho= [q_1,\dots,q_r]$
of $L=$ SO$(d+1)$ are labeled a set of integers subject to
\begin{equation}\label{eq:SOd1rep}
 \varrho = [q_1, \dots, q_r] \quad , \quad  0 \leq q_1 \leq q_2
\leq \dots \leq q_r\ .
\end{equation}
Finite dimensional representations of the rotation group $R =$ SO$(d)$, on the other hand
are labeled by an r-tuple $\tilde \mu = [l_1, \dots, l_r]$ of integers $l_i$ satisfying
\begin{equation}\label{eq:SOdrep}
 \tilde \mu = [l_1, \dots, l_r] \quad , \quad |l_1| \leq l_2 \leq \dots \leq l_r \ .
\end{equation}
Note that the first label $l_1$ may be negative. The representations $\tilde \mu$ with
label $\tilde \mu_l = [0,0,\dots,0,l]$ correspond to symmetric traceless tensors.
Upon restriction from the maximally compact subgroup $L$ to the rotation group $R$,
an irreducible representation $\tilde \mu$ of $R$ can appear in the decomposition of
$\varrho = [q_1, \dots, q_r]$ provided that
\begin{equation}
\label{eq:set}
\tilde \mu \in  \mathcal{J}_{\varrho} \equiv \{\, [l_1,\dots,l_r]\, |\, \ -q_1 \leq
l_1 \leq q_1 \leq l_2 \leq \dots \leq l_r \leq q_r \, \}\ .
\end{equation}
Equipped with all these technical details on the representation theory of orthogonal
groups we are ready harvest some results on the principal series representations.

As a first simple consequence we can describe the decomposition of the
carrier space $\Gamma^{\Delta,\mu}_{S^d}$ into irreducible representations of the
symmetry group $L =$ SO$(d+1)$. The result is
\begin{equation} \label{eq:Gammadec}
  \Gamma^{\Delta,\mu}_{S^d} \equiv \bigoplus_{\varrho | \mu \in \mathcal{J}_\varrho}
   \varrho\ .
\end{equation}
Note that all labels $q_i$ with $i < r$ are restricted to a finite set. On
the other hand, the last label $q_r$ is not bounded from above so that the
decomposition is infinite as it has to be since we build an infinite dimensional
space of sections in a vector bundle over the sphere in terms of finite
dimensional representations of the isometry group $L$.

For us it will be more important to understand which representation $\tilde \mu$
of the rotation group $R = $ SO$(d)$ appears when we restrict the action of $L =$
SO$(d+1)$ on $\Gamma^{\Delta,\mu}_{S^d}$ to the subgroup $R$. Indirectly, the answer is
obtained by combining the decomposition \eqref{eq:Gammadec} with the statement on
the decomposition of irreducible representations $\varrho$ before eq.\ \eqref{eq:set}. But we can
also phrase the result a little more directly. To this end let us write the label
$\mu$ of the $R=$ SO$(d)$ representation that defines the bundle on $S^d$ as
$\mu = [k_1, \dots, k_{r-1},k_r]$. The entries $k_i$ are subject to the
restrictions described in eq.\ \eqref{eq:SOdrep}. Then an SO$(d)$ representation
$\tilde \mu = [l_1, \dots, l_r]$ appears in the decomposition of $\Gamma^{\Delta,\mu}_{S^d}$
provided that its labels obey
\begin{equation} \label{eq:condtmumu}
 -k_1 \leq l_1 \leq k_2 \quad , \quad k_{i-1} \leq l_i \leq k_{i+1} \quad ,
\quad k_{r-1} \leq l_r \ .
\end{equation}
Whenever a representation $\tilde \mu$ can appear, it appears with infinite
multiplicity. As an example let us see how to construct scalar representations
of the rotations group SO$(d)$. This means we are looking for spaces $\Gamma^{\mu}_{S^d}$
that contain a trivial representation $\tilde\mu = (l_1= 0,\dots, l_r = 0)$. According to
the previous conditions, such a representation $\tilde \mu$ can only appear if $\mu =
[k_1, \dots, k_{r-1},k_r]$ takes the form $\mu = [0,0, \dots,0, l]$. The integer $0 \leq l$
remains free. In other words, for a scalar representation to appear, the bundle $\Gamma^\mu$
must be associated with a symmetric traceless tensor representation $\mu = \mu_l$.

\bibliographystyle{JHEP.bst}

\bibliographystyle{plain}
\bibliography{literatureHA}

\printindex

\end{document}